\documentclass[twocolumn,pre,amsmath,amssymb]{revtex4}
\usepackage{graphicx}
\usepackage{dcolumn}
\usepackage{bm}
\usepackage{color}

\newcommand{\ie}{{\it i.e. }}
\newcommand{\eg}{{\it e.g. }}
\newcommand{\etal}{{\it et al. }}
\newcommand{\etalns}{{\it et al.}}

\begin{document}
\title{Nanofluidics, from bulk to interfaces}
\begin{titlepage}
{to appear in Chem. Soc. Rev. (2009)}
\end{titlepage}

\author{Lyd\'eric Bocquet}
\email{lyderic.bocquet@univ-lyon1.fr}
\affiliation{Laboratoire de Physique de la
Mati\`ere Condens\'ee et des Nanostructures; Universit\'e Lyon 1 and CNRS, UMR 5586, 43
Bvd. du 11 Nov. 1918, 69622 Villeurbanne Cedex, France}
\author{Elisabeth Charlaix}
\affiliation{Laboratoire de Physique de la
Mati\`ere Condens\'ee et des Nanostructures; Universit\'e Lyon 1 and CNRS, UMR 5586, 43
Bvd. du 11 Nov. 1918, 69622 Villeurbanne Cedex, France}
\date{\today}
\begin{abstract}
Nanofluidics has emerged recently in the footsteps of microfluidics, following the quest of scale reduction inherent to nanotechnologies.
By definition, nanofluidics explores transport phenomena of fluids at the nanometer scales. Why is the nanometer scale specific ? What fluid properties
are probed at nanometric scales ? In other words, why 'nanofluidics' deserves its own brand name ? 
In this critical review, we will explore the vast manifold of length scales emerging for the fluid behavior at the nanoscales, as well as 
the associated mechanisms and corresponding applications.
We will in particular explore the interplay between bulk and interface phenomena.  The limit of validity of the continuum approaches will be discussed, 
as well as the numerous surface induced effects occuring at these scales, from hydrodynamic slippage to the various electro-kinetic phenomena
originating from the couplings between hydrodynamics and electrostatics. An enlightening analogy between 
ion transport in nanochannels and transport in doped semi-conductors will be discussed.

\end{abstract}

\maketitle

\section{Nanofluidics, surrounding the frame}


Nanofluidics, the study of fluidic transport at nanometer scales, %
has emerged quite recently in the footsteps of microfluidics. Pushing further the limits of fluidic downsizing
is an attracting stream, in the spirit of scale reduction inherent to all micro- and nano- technologies.
Various reasons may be seen as to motivate these novel developments. First, from the point of view of biotechnological (``lab on a chip'') applications, decreasing the scales considerably increases the sensitivity of analytic techniques, with the ultimate goal of isolating and studying individual macromolecules \cite{Eijkel05,Plecis05}.
But also, from the point of view of fluidic operations, nanometric scales allow to develop new fluidic functionalities, taking explicit benefit of 
the predominance of surfaces. Typical examples involve preconcentration phenomena \cite{Pu04}, the development of nanofluidic transistors \cite{Schasfoort99,Karnik05} or the recently proposed nano-fluidic diodes \cite{Siwy02,Karnik07}. 
But the analogy to micro-electronics is somewhat limited: fluid molecules are not
electrons and the notion of large scale integration for fluidic devices, {\it i.e.} nanofluidics as a way of
increasing the density of fluidic operations on a chip even further, is probably not a pertinent goal to reach for fluidic operations. 

But from a different perspective, nanofluidics also carries the hope that new properties will emerge by taking benefit of the specific phenomena occurring at the smallest scales:
new solutions may be obtained from the scales 
where the behavior of matter departs from the common expectations. The great efficiency of biological nanopores (in terms of permeability or selectivity) is definitely a great motivation to foster research in this direction \cite{Sui01,Holt06}. 
There is indeed a Ç lot of room È for improvements at these smallest 
scales: the example of aquaporins (AQP) is interesting in this context. Aquaporins channels are a key component of many biological
processes \cite{Agre00} and play the role of water filters across biological membranes. 
These channels fullfill the conflicting tasks of being both extremely permeable to water, while extremely 
selective for other species \cite{Sui01}. To give an order of magnitude, the 
permeability of the water channel is typically {\it 3 orders of magnitude larger} than 
what would be expected on the basis of the classical fluid framework for the same pore size
\cite{Walz94,Borgnia01}.  %
A potential answer is that AQP, although a filter for water, is mainly {\it hydrophobic}, {\it i.e.}Ê water repellent~! 
Of course some hydrophilic polar nodes are distributed along the pore, so as to keep water in the
mainly hydrophobic environment.
Quoting the terms of Sui {\it et al.} in Ref. \cite{Sui01}, ``{\it the availability of water-binding
sites at these nodes reduces the energy barrier to water transport
across this predominantly hydrophobic pathway, while the relatively
low number of such sites keeps the degree of solute-pore interaction
to a minimum. In balancing these opposing factors the
aquaporins are able to transport water selectively while optimizing
permeability.}''
Beyond this elementary picture, understanding how AQP fullfills its challenging properties would definitely be a source of inspiration and
open new perspectives for technological breakthrough in filtration, desalination, power conversion, ... 
However reproducing such a delicate composite (patched) architecture  in Ç bio-mimetic È membranes
is a great challenge, which requires breakthrough in the conception of its 
elementary constituents. But it points out that surfaces and their chemical engineering are a key actor to optimize fluid
properties at nano-scale. 





However, 
in its roots, nanofluidics is not a new field: as judiciously recalled by Eijkel and van den Berg in their pioneering review on the subject \cite{Eijkel05},
many 'old' fields of physics, chemistry and biology 
already involves the behavior of fluids at the nanoscale. Like Monsieur Jourdain in ``Le bourgeois gentilhomme'' by Moli\`ere, one has done 'nanofluidics' for more than forty years without knowing it \footnote{``Par ma foi ! il y a plus de quarante ans que je dis de la prose sans que j'en susse rien, et je vous suis le plus oblig\'e du monde de m'avoir appris cela.'', which translates into ``By my faith! For more than forty years I have
been speaking prose without knowing anything about it, and I am
much obliged to you for having taught me that.''}. One may cite for example the domains of electro-kinetics (electro-osmosis or -phoresis, ...) with applications in chemistry and soil science, membrane science (ultra-filtration, reverse osmosis, fuel cells, ...), colloid chemistry, and of course physiology and the study of biological channels \cite{Eijkel05}. An interesting question is accordingly wether -- on the basis of the novel 'nanofluidic' point of view -- one may
go beyond the traditional knowledge in those 'old' fields and obtain unforeseen results, {\it e.g.} allowing for better optimization of existing technologies ? Our belief is that the answer to this question is already positive and this is one of the key aspects that we shall discuss in this review.

Finally it is also important to note that 
nanofluidics has emerged recently as a scientific field ({\it i.e.} naming a field as "nanofluidics") also because of the considerable progress made over the last two decades in developping nano-fabrication technologies, now allowing to fabricate specifically designed nanofluidic devices, as well as the great development of new instruments and tools which give the possibility to investigate fluid behavior at the nanometer scale. 
One may cite for example: new electrical detection techniques, Surface Force Apparatus (SFA), Atomic Force Microscopy (AFM), nano-Particle Image Velocimetry (nano-PIV) coupling PIV to TIRF set-up (Total Internal Reflection Fluorescence), as well as the considerable progress made in computational techniques, like Molecular Dynamics simulations.
It is now possible to {\it control/design} what is occuring at these scales, and {\it observe/measure} its effects. This is the novelty of the field, and the reason why nanofluidics now deserves its own terminology. 

The paper is organized as follows:
In the second section we will replace nanofluidics in the perspective of the various length scales at play in fluid dynamics. We shall in particular discuss
the limits of validity of continuum ({\it e.g.} hydrodynamic) descriptions.
In the third section we discuss the dynamics of fluid at interfaces and the nanofluidic tools which have been developped recently to investigate it.
In the fourth section we will explore various transport phenomena occuring in diffuse layers.  
In the fifth section we raise the question of thermal noise in nanofluidic transport.
Finally we will conclude by exploring some general consideration
and expectations about nanofluidics, especially in terms of energy conversion and desalination.

As a final remark, this review, like any review, is our subjective and personal view of the field of nanofluidics and the perspectives one may foresee.
We organized our exploration around the length scales underlying fluid dynamics at the nanometric scales and how nanofluidics allows
to probe the corresponding mechanisms. 
Accordingly, our aim is not to explore exhaustively the -- already large -- litterature of the domain, but merely to disentangle the various effects 
and length scales underlying the behavior of fluids at the nanometer scales.
In doing so, we certainly hope that this review will raise new questions, open new directions and attract people in this fascinating domain. 

\section{Limits of validity of continuum descriptions and nanofluidic length scales }
\label{sec:lengths}

The introduction of the terminology ``nanofluidics'' (furthermore to define a specific scientific field)
suggests that something special should occur for the transport of a fluid
when it is confined in a channel of nanometric size. This leads to an immediate question: why should the nanometer length scale have anything specific
for fluidic transport~? 
Nanofluidics ``probe'' the properties of fluids at the nanoscale: so, what does one probe specifically in the nanometer range ?

Actually 
one may separate two different origins for 
finite-size effects associated with nanometer scales: {\it bulk} and {\it surface}Ê finite-size effects. The former, {\it bulk} effects are intimately associated
with the question of validity of the classical continuum framework, in particular the Navier-Stokes equations of hydrodynamics: when
do such descriptions breakdown ? Can one then expect 'exotic' fluid effects associated with the molecular nature of the fluid ?
On the other hand, {\it surface} effects play an increasingly important role as the ``surface to volume ratio'' increases ({\it i.e.} as the 
confinement increases). We already pointed out the importance of such effects on the example of AQP water channels.
As we show below, the surface effects occur at much larger scales than the 'bulk' deviations from continuum expectations. 

\begin{figure}[htb]
\begin{center}
\includegraphics [width=9 cm] {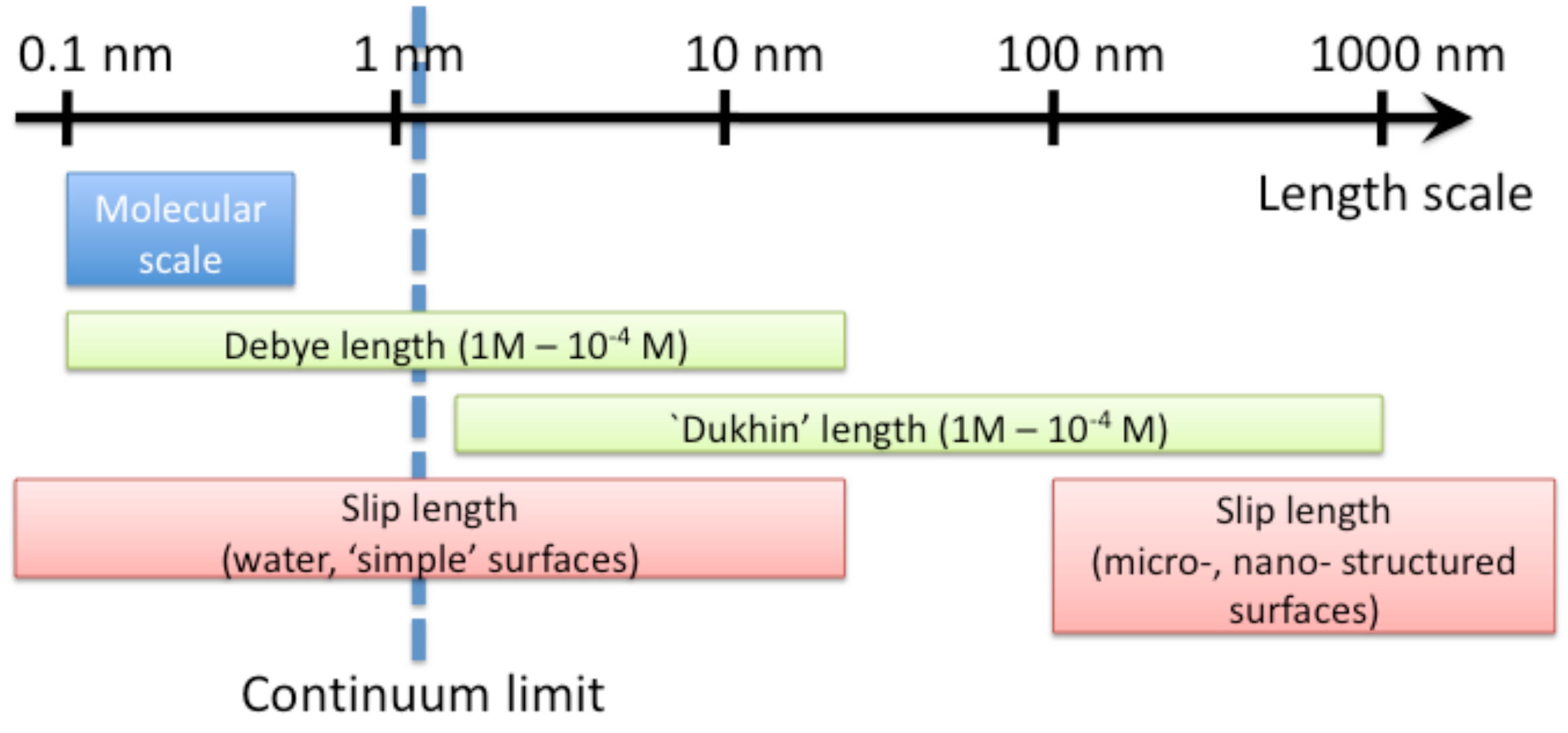}
\end{center}
\caption{Various length scales at play in nanofluidics.}
\label{fig:lengths}
\end{figure}

The discussion on the length scales at play, to be explored in this section, is summarized in Fig. \ref{fig:lengths}.


\subsection{Validity of bulk hydrodynamics}
We first start with a short discussion on the validity of bulk hydrodynamics. In practice, this raises the question: when do the Navier-Stokes (NS) equations break down~?
These equations were developped in the 19$^{th}$ century to describe fluid flows at, say, the human scale.
What comes as a surprise is their incredible robustness when applied to ever smaller scales. 

As a fact, for simple liquids, 
{\bf the continuum framework of hydrodynamics is apparently valid down to the nanometer scale}. In other words, there is no
expected deviations to the bulk NS equations for confinement larger than $\sim 1$nm. This very surprising result is actually suggested by a number
of experimental  and molecular simulations studies. On the experimental side,
one may cite the early work by Chan and Horn \cite{Chan85} and later by Georges {\it et al.} \cite{Georges93} using Surface Force Apparatus, where the 
prediction of hydrodynamics (the Reynolds formula in their case) was verified to be valid for confinement larger than typically ten molecular diameters.
More recently, and specifically for water, 
the works by Klein {\it et al.}Ê \cite{Raviv02}Ê and E. Riedo {\it et al.} \cite{Li07} showed that water keeps its bulk viscosity down to $\sim 1-2$nm, with
a drastic change of behavior for stronger confinements, where the wettability of the confining surface plays a role \cite{Li07}. A similar behavior was found for other liquids like octamethylcyclotetrasiloxane (OMCTS)
\cite{Becker03,Maali06}. 

\begin{figure}[h]
\begin{center}
\includegraphics [width=7 cm] {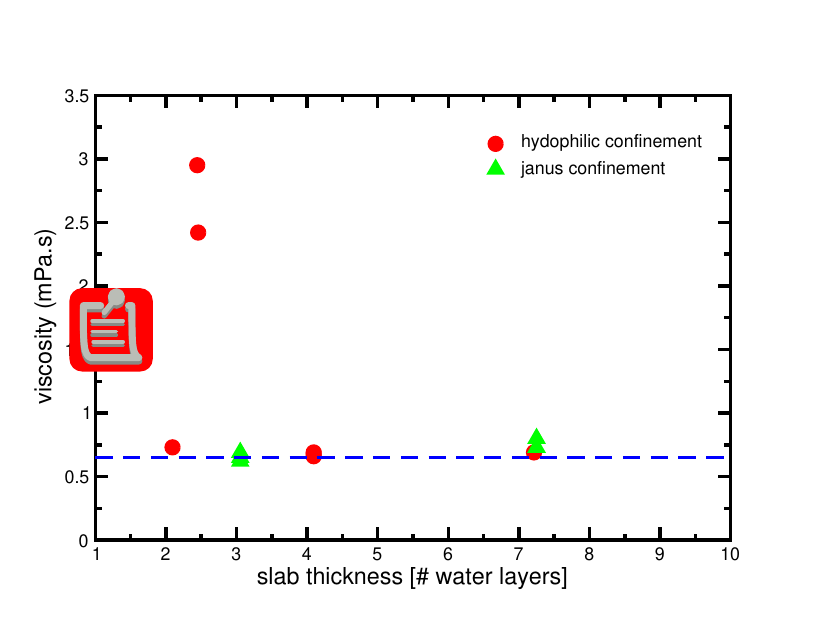}
\end{center}
\caption{Viscosity of water versus confinement, as measured from MD simulations of water confined in a nano-slit made of either two hydrophilic walls, or one hydrophilic and one hydrophobic wall (janus confinement). [courtesy of D. Huang]}
\label{visco}
\end{figure}

This threshold for the applicability of continuum hydrodynamics was also investigated using Molecular Dynamics simulations of confined water, and 
the same value of about $1$ nm
came out of the simulation results \cite{Leng05,Thomas09,Bocquet07}.  
Furthermore, it is interesting to note that beyond the validity of continuum equations, the value of the viscosity also remains quantitatively equal to its bulk
value. 
We show in Fig.\ref{visco} the results of MD simulations for the viscosity $\eta$ of water (SPC-E model) measured in various confinements. 
The value of the viscosity is obtained by measuring the shear-stress on the confining plates for a given shear-rate (corrected for slippage effects).
As shown on this figure, the shear viscosity of water 
keeps its bulk value down to confinements of $\approx 1$nm (typically 3 water layers).

Now, in contrast to the viscosity, other transport coefficients may be more strongly affected by confinement. This is the
case for example of the (self-)diffusion coefficient which was shown to depend algebraically on the confinement width \cite{Joly06,Saugey05}.
As a consequence, the diffusion coefficient in confinement strongly departs from its Stokes-Einstein prediction, $D=k_BT/(3\pi\eta \sigma)$, with $\sigma$
the molecule diameter.
The latter is accordingly not a correct measure of the viscosity, as sometimes assumed \cite{Thomas08,Thomas09}.

Fundamentally the validity of NS equations down to typically $1$nm (for water) is {\it a priori} unexpected.
Navier-Stokes (NS) equations have been developped to account for the fluid dynamics at 'large' scales and
they rely on an assumption of a local and linear relationship between stress and velocity gradients. 

To our knowledge, there is no firmly grounded argument for this validity and we propose here a few tentative leads. 
NS equations, like any 
continuum framework, rely actually on the key assumption of a separation of length- and time- scales between the investigated length
scale and the  'molecular' dynamics. This is the ``hydrodynamic limit'', in which the continuum framework should hold. 
Under this time- and length- scale separation, the
microscopic dynamics associated with a huge number of degrees
of freedom ${\cal N}\approx 10^{23}$ reduces to equations with just a few degrees of freedom (velocity field, pressure, density, temperature, etc.),
while all the complexity is hidden in just a few phenomenological coefficients.  Achieving this averaging out of the "fast variables" is a huge challenge which can be achieved systematically only in a few limiting cases, see {\it e.g.} \cite{Bocquet97} for an explicit example. 
In general, the elimination of fast variables 
is summarized in the Green-Kubo relationship
for the transport coefficients: for example the shear viscosity $\eta$ can be expressed as the integral of the stress-stress correlation function according to
\begin{equation}
\eta = {1 \over V k_BT}Ê\int_0^\infty \langle \sigma_{xy}(t)\sigma_{xy}(0)\rangle_{\rm equ}\, dt
\label{eta}
\end{equation}
with $\sigma_{xy}$ a non diagonal component of the stress tensor \cite{Helfand60} and $\langle\cdot\rangle_{\rm equ}$ denotes an equilibrium average.

The time-scale separation underlying the validity of Eq. (\ref{eta}) requires the microscopic time scales, $\tau_\sigma$, characterizing the stress-stress correlation function, to be much smaller than a hydrodynamic time scale. This time scale is \eg the relaxation time of momentum, which for a given wave vector $q$ has
the expression $\tau_q=(\nu q^2)^{-1}$, with $\nu=\eta/\rho$ the kinematic viscosity and $\rho$ the mass density.
An implicit condition of validity is accordingly
$\nu q^2 \tau_\sigma <1 $.
This fixes the limit for time-scale separation at confinements $w$ larger than a viscous length scale:
\begin{equation}
w>\ell_c=\sqrt{\nu\cdot \tau_\sigma}
\end{equation}
Putting numbers for water,
a typical correlation time for the stress-stress correlation function is in the picosecond range $\tau_\sigma \sim 10^{-12}$s, while $\nu=10^{-6}$ m$^2$.s$^{-1}$. Altogether  this gives for water
\begin{equation}
\ell_c \approx 1 {\rm nm}
\end{equation}
A nanometric characteristic length scale thus emerges naturally as a lower bound for the validity of the notion of viscosity, and thus for the application of standard NS hydrodynamics. 
 This is indeed the experimental value below which strong departure from the continuum framework is observed.
 
 Of course, it would be interesting to explore further the physical contents of this condition: what occurs for smaller confinements ? Also, can one tune  $\ell_c$ in specific systems ? etc. 
 
In the extreme limit of single file transport, the transport behavior is indeed expected to strongly deviate from the bulk expectations, 
with the predicted occurence of anomalous diffusion, and a fast stochastic transport, to cite some examples \cite{Hummer01,Chou98,Chou99}. Apart from biological systems --where molecular pores are the rule --, fabricating channels with molecular size is however still out of reach with state-of-the-art nanofabrication technologies.

{
As a side remark, it would be interesting to extend this discussion to other transport phenomena, like for example heat transfer at nanoscales.  
Heat transfer from nanoparticles and nanostructures to a fluid is an actively growing field, \eg in the context of cooling enhancement by structured surfaces or local heating of fluids via nanoparticles \cite{Merabia09}. Similar arguments as above may be proposed, 
and lead to an equivalent sub-nanometer heat length scale (for water at room temperature). This suggests that the continuum phenomenological picture for heat transport in fluids, involving Fourier law, is expected to hold down to nanoscales, in agreement with observations \cite{Merabia09}.
However a  full investigation of the limit of applicability of the corresponding continuum phenomenological laws remain to be developped in general.}

A conclusion of the above discussion is anyway that for most nanofluidic applications involving water in supra-nanometric confinements, bulk NS equations can be safely used to account for the fluid transport.



\subsection{A broad spectrum of length scales}
Aside from $\ell_c$ introduced above, there is a number of length scales which enter nanofluidic transport. These length scales are all related to surface effects in one way or another. We review
in this section a panel of these length scales, 
climbing up from the smallest -- molecular -- scales to the largest, but still nanometric, scales. This panel is not exhaustive,
and we merely insist on the main length scales which appear generically in nanofluidic problems.
Our aim is to show that there are a number of pertinent scales which lie indeed in the nanometric range. This implies that 
specific nanofluidic phenomena will show up when the confinement compares with these values.

\subsubsection{Molecular length scales}
At the smallest scales, the granularity of the fluid and its components (solvent, ions, dissolved species, ...) should play a role. This is defined by the molecular scale, associated with the diameter of the molecules $\sigma$,  typically in the angstr\"om scale,  $3\AA$ for water. 
As we discussed above, this is the scale where the validity of hydrodynamics breaks down, typically for confinements of 3-5 molecular diameter. 
Various phenomena then come into play.
First,  the structuration and ordering of the fluid at the confining walls plays a role. Such ordering was indeed shown experimentally
to induce oscillatory dissipation in liquid films with a width of several molecular diameter \cite{Becker03,Maali06}.

Similarly non-local rheological properties have been predicted in strongly confined situations  \cite{Hansen07}, associated with a length 
characterizing non-locality typically in a few diameters range (indeed in line with the above limit of NS hydrodynamics). This result echoes recent findings of non-local flow curves in 
confined soft glasses  \cite{Goyon08}. An interesting consequence of non-locality, as pointed out in \cite{Goyon08}, is that
the specific nature of surfaces does influence the global flow behavior in confinement, in line with recent observations in AFM measurements by
Riedo \etal for strongly confined water (less than 1nm) films at a hydrophilic and hydrophobic surfaces \cite{Li07}.
%


But even more drastic are the deviations occuring in the extreme limit where only one molecule can enter the confining pore. 
Very strong correlations and collective motion builds up in the liquid dynamics, leading to the so-called ``single-file'' transport \cite{Levitt73,Rasaiah08}. 
The associated transport differs strongly from the bulk hydrodynamic predictions  and various specific phenomena show up, 
like non-fickian transport \cite{Levitt73}, fast water transport and ion transfer, stochastic flow with the occurence of ``bursts'', etc., see \eg \cite{Rasaiah08}.
Simulations of water transport through single-file
carbon nanotubes (where only one water molecular can pass through the channel) showed indeed flow rate much larger than those predicted from the Poiseuille law of hydrodynamics \cite{Hummer01}.

Note however that this behavior is limited to nanochannels where molecules cannot cross each other, \ie with diameters not exceeding the water molecular length scale, 
like in biological channels. Up to now, artificially produced nanochannels do not reach this limit.
This point is interesting to discuss in the context of recent experimental results obtained for flow through carbon nanotube membrane \cite{Majumder05,Holt06, Whitby08}.
A recent experimental first was achieved in the group of O. Bakajin, who investigated the flow through a membrane constituted of carbon nanotube with a size in the range 1.3-2 nm \cite{Holt06}, following the work on larger nanotubes ($\sim 7$nm) by Majumder \etal \cite{Majumder05}. These works demonstrated massively enhanced flow permeability, as compared
to bulk predictions (in term of slippage, to be discussed below, this would correspond to slip lengths up the micron, far above any expected result for these surfaces).
More recent experiments performed with membranes made of wider carbon nanotubes suggest a large, but still smaller, enhancement of the permeability \cite{Whitby08}.
Therefore, while it would be tempting to discuss these results in the context of single-file transport and the predictions
by Hummer \etal on fast single-file transport in sub-nanometer carbon nanotube \cite{Rasaiah08}, this point of view is difficult to justify: as discussed above, fluid
flow in nanotube pipes with a supra-nanometric diameter  can be safely described by continuum approaches (Stokes equation) and the dynamics is definitely not single file.
Molecular Dynamics simulations of flow through similar carbon nanopipes indeed suggest large flow enhancement \cite{Joseph08b,Thomas08}, but still far below the 
experimental results. There is therefore a strong need to perform flow experiments through a single object, \ie through a single carbon nanotube. This is still an experimental
challenge.

Altogether transport at the molecular scale offers a rich panel of transport behavior. This is definitely the scale where, 
as we quoted in the introduction, new solutions and technological breakthrough could emerge as the behavior of matter departs from the common expectations.
This suggests that the molecular scale would be the ultimate scale to reach for nanofluidics. But at present, producing channels of sub-nanometric size 
remains a technological challenge.
%

\subsubsection{Electrostatic length scales and their dynamic influence}
\label{sec:electrostat}

The scales associated with the interaction of the flowing matter with its environement also enter the game. And of course, electrostatic
plays a key role in this context. We leave aside interactions deriving indirectly from electrostatics
to focus on the interactions between charged species. These obviously concern ions dissolved in the fluid: ion-ion interaction, and interactions of
ions with charged surfaces. This also involves the dipolar interaction among water molecules, as well as H-bonding. 

Electrostatics involves a rich panel of length scales. And more importantly, due to the long range nature of electrostatics, these length scales are intimately coupled: 
effects occuring at the smallest electrostatic lengths do climb the scales to span the whole range.

\vskip0.5cm
{\it a. Bjerrum length (and some derivatives)}\\

We start with the 
Bjerrum length, $\ell_B$, which is defined as the distance at which the electrostatic interaction $ {\cal V}_{\rm el}$ between two charged species
becomes of the order of the thermal energy, $k_BT$: $ {\cal V}_{\rm el}(\ell_B) \approx k_BT$. For two ions, with valence
$z$, embeded in a dielectric medium with dielectric constant $\epsilon $, this takes the form
\begin{equation}
\ell_B={z^2 e^2 \over {4\pi \epsilon  k_BT}}
\end{equation}
($e$ the elementary charge).  For bulk water at ambiant temperature and a valency $z=1$,  this gives $\ell_B=0.7$nm. The Bjerrum length therefore compares with the molecular range in this
 case. Note however that for multi-valent ions or for organic solvent with lower
dielectric constants, $\ell_B$ may be much larger and thus separate from the molecular range.

Other length scales could be introduced along the same lines by comparing thermal energy to \eg charge-dipole, dipole-dipole, etc. interactions, 
taking into account the multi-pole interaction under consideration. This leads {\it a priori} to length
scales smaller than the Bjerrum length introduced above. For the charge-dipole interaction, this gives for example
$\ell_d=\left(\ell_B {p\over e}\right)^{1/2}$, with $p$ the dipole strength: for water  parameters, $\ell_d$ is in the Angstr\"om range. 
It does correspond to the typical thickness of a hydration layer around an ion \cite{Israelachvili96}.

By definition, the Bjerrum length (and its derivatives like $\ell_d$ above) is the scale below which direct electrostatic interactions dominate over thermal effects. 
Its consequences on nanofluidic transport is thus expected at molecular scale. For example,
for confinement below $\ell_B$ one expects a large free-energy cost to undress an ion from its hydration layer and make it enter a molecular pore.  
This has therefore immediate consequences on the filtering process of charged species, as in biological ion channels.

But as pointed out above, a strong interplay with upper scales is expected due to the long range nature of electrostatic interactions. 
For example, as we will discuss in section \ref{sec:DL}, ion-specific effects at interfaces, the origin of which occurs at a scale $\ell_B$, 
have strong effects on the electro-kinetics at nanometric, micrometric and even larger scales. 



\vskip0.5cm
{\it b. Debye length scale} \\Ê

A central concept of electrostatics is the notion of electrostatic diffuse layer (EDL).
 At a charged interface, the EDL is the region where the surface charge is balanced by the cloud of counterions. In this region the ion concentration profiles depart from their bulk values due to the interaction of ions with
 the surface charge. This is therefore the region where local electro-neutrality is not obeyed.

The Debye length emerges naturally from the Poisson-Boltzmann theory and characterizes the electrostatic screening in the bulk electrolyte  
\cite{Andelman95}. 
It is defined in terms of the salt concentration $\rho_s$ according to the expression:
\begin{equation}
\lambda_D= \left(8\pi \ell_B \rho_s\right)^{-1/2}
\label{Dlength}
\end{equation}
Note that in this expression, as in the following, $\rho_s$ denote the salt concentration {\it in the reservoirs}, thereby fixed by the chemical potential of the ions.
The value of the Debye length depends on the salt concentration and ranges typically between tens of nanometer (30 nm for $\rho_s=10^{-4}$M) down to Angstr\"oms (3 \AA \, for $\rho_s=1$M). 
In physical terms,  the electrostatic free-energy of an ion dressed by its spherical cloud of counterions with size $\lambda_D$ is of the order $k_BT$.

The Debye length characterizes the width of the EDL. 
Interestingly, it is independent of the surface charge and only depend on the
bulk ion concentration. 
However,  in the limiting case of a salt free solution (with $\lambda_D \rightarrow \infty$), the width of the diffuse layer is merely fixed by the
so-called Gouy-Chapman length, to be defined in the next paragraph. 
 
The Debye length plays a central role in nanofluidics for various reasons. First, as quoted above, this is the region in the fluid close to charged surfaces
where local charge electroneutrality is broken. Under an applied electric field, this is thus the region where volume electric forces, $f_e=\rho_e E_e$, will apply, with 
$\rho_e$ the charge density and $E_e$ an applied electric field. On the other hand these forces vanish in the bulk of the
material due to a vanishing charge density  $\rho_e\rightarrow 0$ far from the surfaces.
Therefore tuning the fluid properties inside the nanometric EDL is expected to affect the whole response of the system under an electric field. Accordingly, 
tuning the dynamics of the fluid in the nanometric EDL, or modifying its structure, will have a macroscopic impact  on the
fluid dynamics at scales much larger than the Debye length. We will come back extensively on this question below,
in section \ref{sec:EOslip}.

Second, a specific behavior is expected when the Debye layers overlap in a nanopore, which occurs when its size is of the order of twice the Debye length. This phenomenon 
has a strong effect on fluidic transport and has been the object of intense research recently, as reviewed recently
by Schoch \etal \cite{Schoch08}, see Sec. \ref{sec:DL}. One may cite in particular the phenomena of ion enrichement and exclusion \cite{Plecis05}. While these phenomena were known for a long while in membrane technology, they found new applications in the field of nanofluidics. It is at the root of  a number of novel fluidic phenomena, such as permselectivity \cite{Plecis05}, nanofluidic diodes \cite{Siwy02,Karnik07} or surface dominated ion transport \cite{Stein04}.

A key practical aspect is that the condition of Debye layer overlap is much less stringent to fullfill with actual nanofabrication techniques. These are now
able to produce individual nanopores with at least one dimension in the range of ten nanometers, 
\ie potentially suited for Debye layer overlap. 

\vskip0.5cm
{\it c. Length scales associated with surface charge: Gouy-Chapman and Dukhin  } \\Ê

Other electrostatic length scales may be constructed on the basis of surface electrostatic properties.
We already mentioned the {\it Gouy-Chapman length scale}, which shows up for the behavior of a salt solution very close to a charged surface. For a surface charge density
$\Sigma\, e$, this has an expression $\ell_{GC}=1/(2\pi \Sigma \ell_B)$. In contrast to the Debye length, 
the Gouy-Chapman length 
depends explicitly on the surface density $\Sigma$ of charges
 of the confining surface ($\Sigma$ in units of $m^{-2}$), but not on the bulk ion concentration.
Physically, the  Gouy-Chapman length can be defined as the distance from the wall where the electrostatic interaction of a single ion with the wall becomes of the order of the thermal energy.
For typical surface charges, say $\Sigma e \sim$ 50 mC/m$^2$ ($\approx 0.3\, e/$nm$^2$) to fix ideas, then
$\ell_{GC} \sim 1$nm.
The Gouy-Chapmann length plays a role for solutions with very small salt concentration. This has an influence on the conductance in nanochannels as we discuss in Sec.~\ref{sec:surfcond}.

More interesting is a length which can be defined on the basis of the comparison between the bulk to the surface electric conductance (relating 
electric current to an applied electric field). This introduces what can be termed as a ``Dukhin length'', by analogy to the Dukhin number
usually introduced for colloids  \footnote{We specifically thank D. Stein for very interesting discussions concerning this point.}. 
Indeed the conduction probes the number of free carriers (ions), so that
in a channel of width $h$ and surface charge $\Sigma$,  the equivalent bulk concentration of
counterions is $2\Sigma/h$. One may define a Dukhin number
$Du= \vert\Sigma \vert/ (h \rho_s)$, where $\rho_s$ is  the concentration of the salt reservoir  \cite{Hunter}. 
A `Dukhin length' can then be defined as 
\begin{equation}
\ell_{Du}= {\vert\Sigma\vert\over \rho_s}.
\end{equation} 
To put numbers, for a surface with a 
surface charge density $e\Sigma\sim 50 $mC/m$^2$ ($\approx 0.3 e/$nm$^2$), $\ell_{Du}$ is typically 0.5 nm for $\rho_s=1$M, while $\ell_{Du}=5$ $\mu$m for $\rho_s=10^{-4}$M !
The Dukhin length characterizes the channel scale below which surface conduction dominates over the bulk one.
In a different context of charge discontinuities at surfaces, it has also been interpreted in terms of an electrokinetic ``healing'' length 
\cite{Khair08}.

This length scale can actually be rewritten in terms of the Debye and Gouy-Chapman lengths 
as $\ell_{Du} \sim \lambda_D^2/\ell_{GC}$. 
Note also that in the limit where the Debye length is large compared to the Gouy-Chapmann length, $\ell_{GC} < \lambda_D$, the non-linear Poisson-Boltzmann expression for the electrostatic potential
\cite{Andelman95}
allows to rewrite the above length as $\ell_{Du}\sim \lambda_D \exp[ e \vert V_s\vert/2k_BT]$, with $V_s$ the surface potential. The corresponding form for the Dukhin number is more standard in colloid science \cite{Hunter}.

This length plays an important role for the conductance in nanochannels, cf Sec.\ref{sec:surfcond}, where 
surface effects are shown to strongly affect conductance  \cite{Stein04,Karnik05b}.

\subsubsection{Slip lengths and surface friction}

Up to now we merely considered length scales characterizing the structure of the fluid and its components (ions, ...). However the {\it dynamics}Ê  of 
fluids at interfaces introduce various length scales. This is in particular the case of the so-called {\it slip length}, $b$, characterizing the hydrodynamic boundary
condition of a fluid at its confining interfaces. The latter is defined according to the Navier boundary condition (BC) as \cite{Bocquet07}
\begin{equation}
b\, \partial_n v_t = v_t
\label{slip}
\end{equation}
with $n,t$ points to the normal and tangential directions of the surface, $v_t$ is the tangential velocity field and $b$ the slip length.
The slip length characterizes the friction of the fluid at the interface and large slip lengths are associated with low liquid-solid friction.

This point has been amply explored experimentally and theoretically: a key result which emerges from this measurements is that
the slip length of water at solid surfaces depends crucially on the wettability of the surface \cite{Bocquet07,Huang08b}.
We will come back  more exhaustively on this point in Sec.~\ref{sec:slip}. At this level, one should keep in mind that
slip lengths in the range of a few ten
nanometers are typically measured on hydrophobic surfaces, while $b$ is sub-nanometric on hydrophilic surfaces. 
Note however that very large slip length, in the
micron range, may be obtained on nano- and micro- structured interfaces \cite{Joseph06,Lee08}.

This offers the possibility to modify the nanofluidics in pores using chemical engineering of the surfaces. As the pore size compares with $b$
a considerable enhancement of fluid transport is accordingly expected. Furthermore one may remark that these values of $b$ also compares with typical Debye
lengths. Therefore slippage effects are expected to affect ion transport at charged surfaces. 
We will discuss these aspects more exhaustively below.

A final remark is that other surface related lengths may be constructed. This is the case in particular of the Kapitza length, which characterizes the boundary condition
for thermal transport across an interface, see \eg  \cite{Bocquet07}. The Kaptiza length behaves quite similarly to the slip length and strongly depends on the
wettability of the interface \cite{Ge06}. 


\subsubsection{Other length scales}

The review of length scales above is not exhaustive and many other length scales may enter the game, depending on the problem, geometry and system
considered. 
For example, in the context of transport of macromolecules (colloids, polymers, DNA, RNA, proteins,  ...), the typical size of the particle -- diameter or radius or gyration --  plays of course a central role.
More precisely a key quantity is the free energy associated with the confinement of the macromolecule, ${\cal F}_c$ which fixes the partitioning of the molecule
in the nanopore w.r.t. the bulk. The probability of passage of a macromolecule through a pore is expected to scale like $\exp[- {\cal F}_c/k_BT]$, hence fixing its permeability \cite{deGennes}.
For example for a polymer chain confined in a pore with size $D$, the entropic cost of confinement takes the form ${\cal F}_c\approx k_BT (R_g/D)^{1/\nu}$,
with $R_g=a\times N^{\nu}$ the radius of gyration of the polymer ($a$ the monomer size and $N$ the polymer length). For a polymer in good solution,
the Flory exponent is $\nu=5/3$. Pores the size of which is below the radius of gyration of the molecule thus act as molecular sieve. This has been explored in 
various experiments of polymer translocation in nanopore following the pioneering work of Bezrukov \etal \cite{Bezrukov93,Bezrukov96} and 
KasianowiczÊ\cite{Kasianowicz96}, as well
as in the context of DNA separation \cite{Han00}.


\subsection{Some practical conclusions}

The discussion above is summarized in Fig.\ref{fig:lengths}, where we organized the various scales at play along the scale axis. 
This points to several conclusions.
\begin{enumerate}
\item A number of specific phenomena occur in the nanometric range, which indeed justifies the specificity of the dynamics of fluids in the nanometer range, \ie nanofluidics.

\item An important conclusion is : For water under normal conditions, the Navier-Stokes equation 
remains valid in nano-channels down to typically 1-2 nanometers.
This is good and bad news. Good because {\it one may safely use NS equations for most nanofluidic phenomena}. This robustness of continuum 
hydrodynamics is remarkable.

But this is in some sense {\it bad news} since it means that reaching specific effects associated with the granularity of the fluid requires molecular confinement, \ie
confinement below the nanometer. Producing nanochannels with such sizes is technologically difficult to achieve. 
On the other hand, this probably explains why most biological channels have a molecular entanglement: to reach specificity and avoid the ``universality'' associated
with continuum equations,  one should reach the fluid molecular scale. 

\item This suggests that the molecular pore is the ultimate scale to reach for nanofluidics, where,
as we quoted in the introduction, new solutions and technological breakthrough could emerge as the behavior of matter departs from the common expectations.
But at present, there is however still a long way before this scale is technologically accessible.

\item 
While it remains difficult to tune the bulk behavior of the confined fluid, there are much more room and possibilities to benefit from surface effects.
This is clearly apparent in Fig. \ref{fig:lengths}, where surface effects enter the nanofluidic game at much larger length scales. Surface effects enter
in particular via the Debye length, Dukhin length, slip length, all typically in ten  nanometers range (and even more for the Dukhin length). This suggests that specific effects will show up 
when one of these lengths will compare with the pore width. Furthermore, particular effects  should also occur when two of these lengths 
become comparable, independently of the confinement. This will be explored in Sec. \ref{sec:DL}.

Altogether nanofluidics is a incredible playground to play with surface effects !

\end{enumerate}

\section{Dynamics at surfaces and the toolbox of nanofluidics}

\label{sec:slip}

We have discussed above that the bulk laws of hydrodynamics are valid down to very small
scales, typically $\sim 1$nm, so that bulk Navier-Stokes equations can be used for most
nanofluidic flows. However as the size of the nanochannel decreases, the dynamics
at its surface should play an increasingly important role. Navier-Stokes equations require
boundary conditions (BC) for the hydrodynamic flow at the device's surface, and a specific
knowledge of these BC is a pre-requisite to apprehend flow at the nanoscale.

As we quoted above, the BC at a solid surface introduces a new length, the so-called Navier length or slip length, which relates  the tangential velocity $v_t$  to the shear rate 
at the wall \cite{Navier}:
\begin{equation}
v_t=b  \frac{\partial v_t}{\partial z}
\end{equation}
Obviously the control of the slip length is of major importance for  flows in confined geometries,  and it can have dramatic consequences on the pressure drop, electrical, and diffusive transport through nanochannels. For instance  the relative increase in  hydraulic conductance of a cylinder of radius $r$ due to wall slippage   is $1+6b/r$; it can be much larger for the electrical conductance or electro-osmosis, as we will discuss in the sections below.

As a consequence a sustained interest has been devoted in the last ten years for the 
investigation of the BC and its dependence on interfacial properties, such as  the surface topography and the liquid-solid interactions. Reviews on this subject are  \cite{LaugaBrenner,NetoCraig,Bocquet07}.

\begin{figure}[!htbp]
\includegraphics[width=4cm]{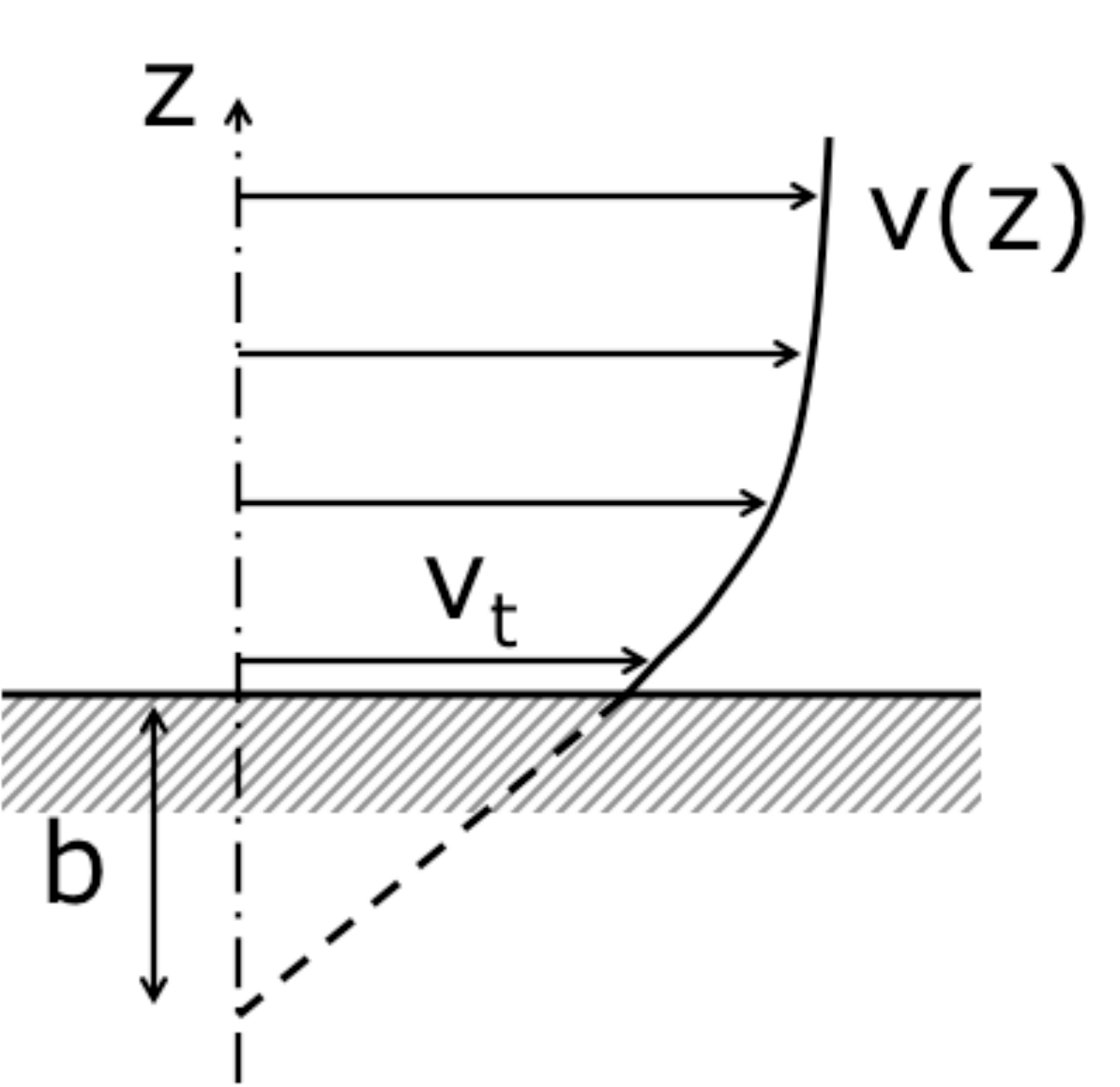}
\caption{Hydrodynamic slip and definition of the slip length $b$: the slip length is the distance in the solid at 
which the linear extrapolation of the velocity profile vanishes.}
\end{figure}

\subsection{Theoretical expectations for slippage}

The theoretical understanding of slippage has been the object of intense research over the last decade, see Ref. \cite{Bocquet07} for a review.
The BC has been studied theoretically by Molecular Dynamics simulations, as well as in the context of linear response theory \cite{Robbins90PRA,BarratBocquetPRL93,BarratBocquetPRE94,KoplikCieplak01,TroianPriezjev04,Huang08b,Sendner09}. It is now   well understood   that the BC on atomically smooth surfaces depends essentially on the structure of the liquid at the interface,  itself  determined by its interactions, its commensurability with the  solid phase,  and its global density  \cite{Bocquet07}. 

In order to connect the slip length to the interfacial properties, it is useful to interprete the slip length in terms of liquid-solid friction at the interface. The friction force at the liquid-solid interface is linear in slip velocity $v_t$:
\begin{equation}
F_f= - {\cal A}\, \lambda \, v_t
\label{friction}
\end{equation}
with  $\lambda$ the liquid-solid friction coefficient and ${\cal A}$ the lateral area. By definition the slip length is related to the
friction coefficient according to $b=\eta/\lambda$, with $\eta$ the bulk viscosity.
As a phenomenological friction coefficient,  $\lambda$ can be expressed in terms of equilibrium properties in the form of a Green-Kubo (GK) relationship
\cite{BarratBocquetPRE94,Barrat99,Bocquet07}:
\begin{equation}
\lambda= {1\over {\cal A}\, k_BT} \int_0^\infty \langle F_f(t)\cdot F_f(0)Ê\rangle_{\rm equ} dt
\end{equation}
where $F_f$ is the total (microscopic) lateral force acting on the surface.
In practice, this expression is difficult to estimate. However some useful information on the slip length may be extracted from it. 

A rough estimate of slippage effects can be proposed  \cite{Sendner09}.
One may indeed approximate the GK expression as $\lambda \approx {1\over {{\cal A} k_BT}} \langle F_f^2Ê\rangle_{\rm equ} \times \tau$,
with $\tau $ a typical relaxation time and $\langle F_f^2Ê\rangle_{\rm equ} $ the mean-squared lateral surface force. 
We then write $\tau \sim \sigma^2/D$ where $\sigma$ is a microscopic characteristic length scale and $D$ the fluid diffusion coefficient,
while the rms force is estimated as $\langle F_f^2Ê\rangle_{\rm equ} \sim C_\perp \rho \sigma (\epsilon/\sigma)^2$, with
$\epsilon$ a typical fluid-solid molecular energy, $\rho$ the fluid density and $C_\perp$ a geometric factor that accounts for
roughness effects at the {\it atomic level} (large $C_\perp$ corresponding to a larger atomic roughness) \cite{Huang08b,Sendner09}. Altogether
this provides a microscopic estimate for the slip length as
\begin{equation}
b \sim {k_BT \eta  D\over {C_\perp\rho \sigma \epsilon^2}}
\label{formuleb}
\end{equation}
A more systematic derivation leads to a very similar result \cite{Barrat99,Bocquet07}, with 
Eq.(\ref{formuleb}) multiplied by  the inverse of the  
structure factor of the fluid, $S_w(q_\parallel)$, computed at a characteristic wave-vector of the solid surface $q_\parallel$:
$b\propto S_w(q_\parallel)^{-1}$.
This term measures a kind of commensurability of the fluid with the underlying solid structure, in full analogy with solid-on-solid friction. Note also that the slip length depends on the product
$\eta \times D$, so that according to the Stokes-Einstein relationship, the slip length is not expected to depend on the bulk fluid viscosity
(except for specific situations where these two quantities can be decorrelated).
The scalings proposed by the above simple estimate are in good agreement with molecular dynamics results \cite{Bocquet07,Sendner09}.

According to Eq.(\ref{formuleb}) significant slippage should occur on very smooth surfaces, in case of a low density at wall - which requires both a moderate pressure and unfavorable liquid-solid interactions -,  or a low value of $S_w(q_{\parallel})$, which characterizes liquid/solid commensurability. It also suggests that low energy surfaces, \ie  with small liquid-solid interaction $\epsilon$, should exhibit large slippage.
Accordingly, hydrophobic surfaces should exhibit larger slip length than hydrophilic ones.

\begin{figure}
\includegraphics[width=6cm]{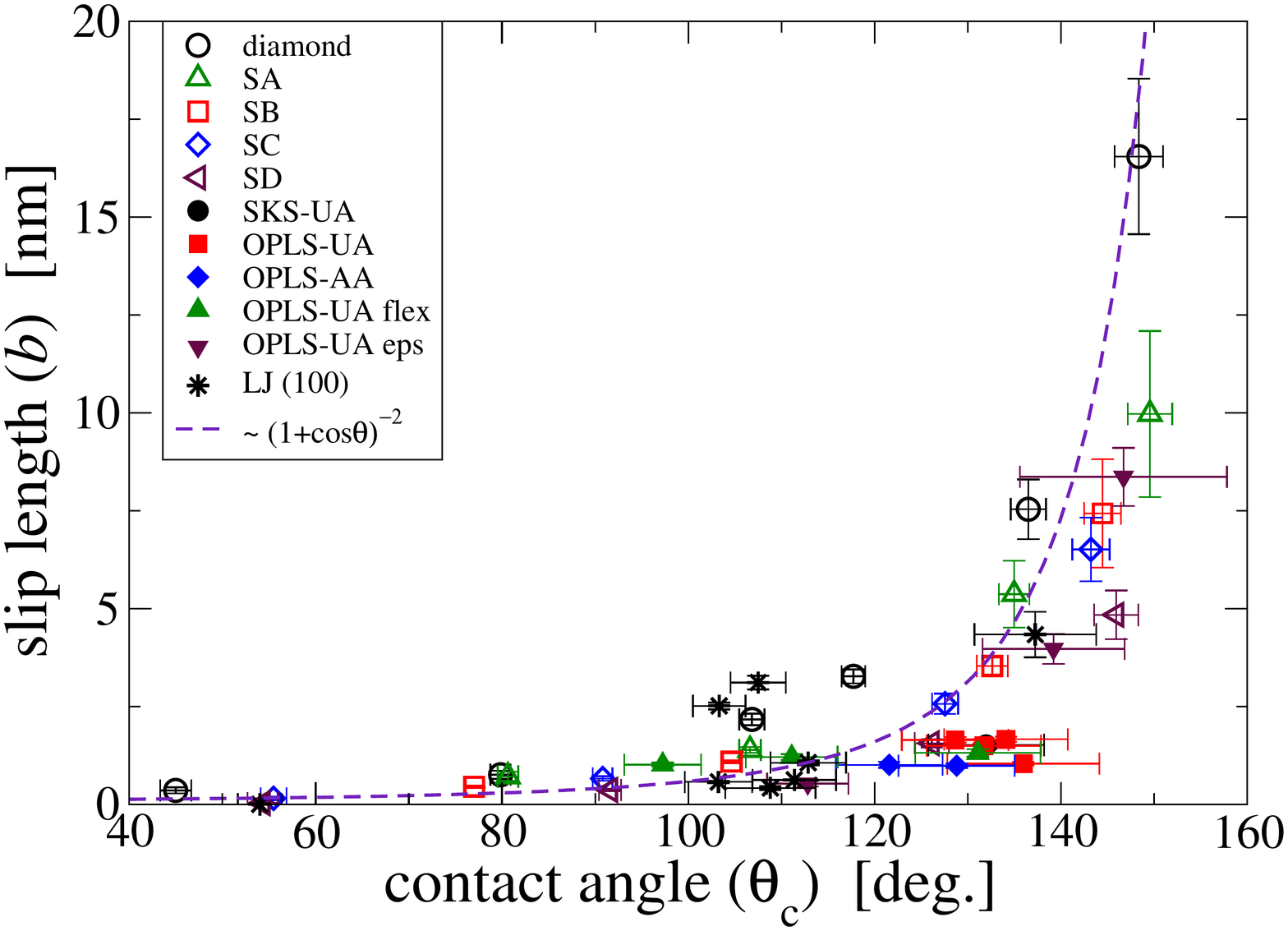}
\includegraphics[width=6cm]{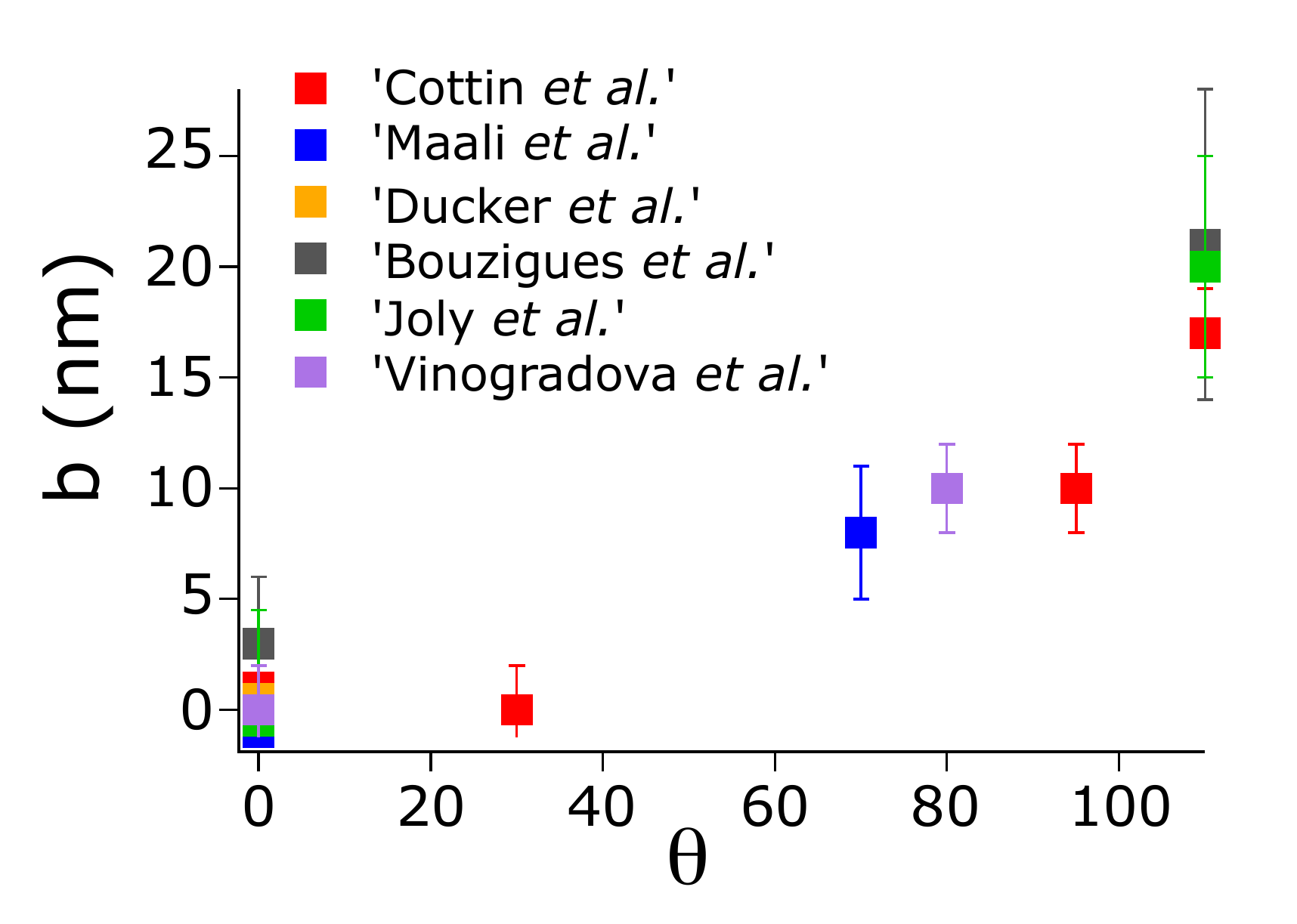}
\caption{Slip length $b$ of water as a function of the contact angle on various smooth surfaces. {\it Top} : Results from Molecular Dynamics simulations by Huang \etal for the slip length of water on various surfaces, from Ref. \cite{Huang08b}. The dashed line is a theoretical fit scaling like $b\propto (1+\cos\theta)^{-2}$. 
{\it Bottom} Compilation of various experimental results. }
\label{compilation}
\end{figure}

We plot in Fig. ~\ref{compilation} (top) results for the slip length obtained by Molecular Dynamics of a water model over a broad variety of surfaces \cite{Huang08b,Sendner09}.
A 'quasi-universal' dependence of the slip length on the contact angle is measured, with -- as expected from the above arguments -- an increase of the slip length for more hydrophobic surfaces.

\subsection{Experimental results}Ê
On the experimental side a race has engaged toward the quantitative characterization of BCs within the nanometer resolution, using the most recent developmments in optics and scanning probe techniques
\cite{LaugaBrenner, NetoCraig}.   This experimental challenge has 
generated considerable progress in the experimental tools for nanofluidic measurements, and  the values of the slip lengths reported had a tendancy to decrease as the resolution and robustness of the techniques has increased.   
An intrinsic difficulty stems from the fact that  the BC is a continuous medium concept involving an hydrodynamic velocity averaged on many molecules in the liquid and extrapolated to the wall: it does not  simply reduce to the velocity close to the surface of molecules or tracer particules, which are furthermore submitted to brownian motion and reversible adsorption. 
Also, the surface has to be smooth and homogeneous on the scale of the probed area in order to characterize an {\it intrinsic} BC, and not an {\it effective} slip length.
Thus, the problem of the BC  has been a discriminating test of the capacity of instruments to perform  quantitative mechanical measurements  at the nanoscale.
We propose here a brief overview of the current  state-of-the-art. This offers the opportunity to quote the various tools which have been developped recently to study flow and fluid dynamics at the nanofluidic scales.

The Surface Force Apparatus provided the first sub-nanometer resolution measurement of BCs,
based on the viscous force $F_v$ acting between two   crossed cylinders whose distance $D$ is measured within the Angstrom  resolution by the so-called FECO interferometric fringes
\cite{IsraNature76}:
\begin{eqnarray}
F_v= \frac{6\pi \eta R^2 \dot D}{D}f^*\\
f^*_s (D/b)= \frac{D}{3b}[(1+\frac{D}{6b}){\rm ln}(1+\frac{6b}{D})-1)]
 \label{formuleOlga}
\end{eqnarray}
with $\eta$ the liquid viscosity and $R$ the  cylinder radius \cite{VinogradovaLA95}.
In case of a no-slip BC, $F_v$ reduces to the Reynolds force
between a sphere and a plane. 
Chan and Horn \cite{Chan85} investigated the flow of various organic liquids confined between atomically flat mica surfaces and 
 showed that they obeyed the no-slip BC, with the no-slip plane located about one molecular size inside the liquid phase. 
 Their results were confirmed by other experiments  running the SFA in a dynamic mode,
and extended to water and to various (wetting) surfaces    \cite{IsraelachviliJCSS86,Georges93}. 
Further investigations focused  on non-wetting surfaces and
  reported the existence of very large slip lengths ($\simeq 1\mu m$) associated to a
 complex behaviour, with slippage appearing only under confinement and above a critical shear stress  \cite{zhugranickPRL01}. However it was  shown that  elasto-hydrodynamics  effects due to non perfectly rigid surfaces (such as glued mica thin films) retrieve this typical behaviour \cite{SteinbergerPRL08}.
 Experiments on rigid surfaces later reported the existence of substantial slippage of water on smooth homogeneous non-wetting surfaces, with a slip length increasing with the contact angle and reaching a range of 20nm at contact angles of 110$^o$ \cite{CottinPRL05, SteinbergerLA08}, see  Fig.~\ref{compilation}.

The AFM with a colloidal probe operates along a similar principle but with a smaller probed area than the SFA. 
In pioneering investigations Craig \etal found  a shear-rate dependent BC of sucrose solutions on partially wetting SAMs, with slip lengths reaching 20nm at high velocities \cite{Craig2001}, while Bonacurso {\it et al.} found a constant (no shear dependent) slip length of 8-9nm for water on wetting mica and glass surfaces \cite{BonaccursoPRL02}. These dissimilar findings also did not agree with available theoretical results. 
However    Ducker \etalns, using an independant optical measurement of the probe-distance in the AFM,
found a regular no-slip BC on a similar system as the one investigated by Craig \etal
\cite{DuckerPRL07}. It was recently shown by Craig \etal that the lateral tilt of the cantilever was a significant issue in interpretating hydrodynamic forces \cite{Craigsub}.
Vinogradova \etal also evidenced  the important effect of the dissipation due to the cantilever \cite{VinogradovaRSI01}.   Using a "snow-man" probe made of two spheres glued on the top of each other, they showed a perfect agreement with the no-slip BC on hydrophilic surfaces, confirming SFA experiments, and a moderate slip length of 10nm of water onto rough hydrophobic surfaces. 
Finally dynamic measurements using an oscillatory drive were implemented in the AFM by Maali \etal so as to measure the dissipation coefficient with high accuracy \cite{MaaliAPL06}.
Using this technique they found a no-slip BC of water onto mica surface, and a 8nm slip length on atomically smooth HOPG graphite with contact angle 70$^o$ \cite{MaaliAPL08}.

An 
alternative way to measure interfacial hydrodynamic dissipation is through the dissipation-fluctuation theorem, for instance by measuring
the modification of the brownian motion of a probe at the vicinity of a surface. In Joly \etal experiment \cite{Joly06} the average transverse diffusion coefficient of colloidal particles confined in a liquid slab was measured   by Fluorescence   Correlation Spectroscopy as a function of the film thickness, and related to the boundary condition on the slab walls. A resolution better than 5nm is obtained with particles of 200nm nominal diameter, by averaging the data over 100000 events crossing the measurement volume. 
The originality of the method is to give access to the boundary condition without any flow, i.e. really at thermodynamic equilibrium.  It is thus directly comparable to theoretical predictions. A collateral result  is the direct probe of a property of great interest in nanofluidics, i.e. the motion of a solute close to a wall.

Besides the dissipative methods, great progress was achieved in the development of optical methods 
 giving direct access to the flow field and its extrapolation to the solid wall with a sub-micrometer resolution.

The Micro-Particle Image Velocity ($\mu$-PIV) technique has been widely used in microfluidics.
First attempts to determine the  BC yielded slip lengths of micrometer amplitude  \cite{trethewayPF02} ,
 difficult to conciliate with theoretical expectations,
although the variation of slippage with the liquid-solid interaction were consistent.
Gaining better resolution raised many issues. 
Besides colloidal forces acting on particles close to a wall, which have to be taken into account, such as DLVO forces or electro-osmosis
induced by the streaming potential,  an important 
source of uncertainty is  the determination of the wall position: locating the latter from the signal of moving particles lead to a systematic bias  because of the depletion induced by the colloidal lift \cite{VinogradovaPRE03}. By using the signal of non-moving particles adsorbed to the wall Joseph et al \cite{Joseph05}  increased the resolution on the slip length to 100nm in a configuration similar to \cite{trethewayPF02}, and found no slippage at this (100nm) scale whatever the wettability of the surface. A recent investigation
using double-focus FCS confirmed the existence of slippage of a few ten nanometers on hydrophobic surfaces \cite{Vinogradova09}.
A major improvement toward nanometric resolution is the Total Internal Reflection Fluorescence method which uses  an evanescent wave  at the interface to be probed,  and allow to measure the distance of a tracer particle to the wall through the exponential decay of the signal amplitude. Bouzigues {\it et al.} found a water slip length of $21$nm ($\pm 12$nm) on OTS coated glass capillaries \cite{Bouzigues08}, while Lasne \etal measured a water slip length of $45$nm ($\pm 15$nm) on a similar surface \cite{Lasne08}. These results are consistent and the different values of slip length may well be ascribed to slight variations in surface preparation.
Note that in such measurements, the size of the probe itself  is a major issue in the race toward nanometric resolution:  Brownian motion increases as the particle size decreases, not only adding noise to the measurement, but allowing particles to leave the focal plane between two successive images which decreases the locality of the measurement. Some specific statistical treatment has to be performed in order to avoid any bias in the estimated velocity \cite{Bouzigues08,Lasne08,Ducker09}.

A summary of the results obtained with water on various smooth surfaces is gathered in Fig.\ref{compilation} plotting the slip length as a function of the contact angle. The  various methods show the same trend: water does not slip  on hydrophilic surfaces, and develops significant slippage   only on strongly hydrophobic surfaces.  The highest slip lengths  reach the range of 20-30nm: accordingly hydrodynamic slippage is non relevant in microfluidic devices,  but  of major importance in nanochannels. 
On this basis one expects that water flow in highly confined hydrophobic pores, such as mesoporous media or biological channels ({\it c.f.}Êthe Aquaporin example in the introduction), should be much less dissipative than ordinary Poiseuille flow.
 The results obtained with the various methods show a very good qualitative agreement with  theory and molecular dynamics simulations performed with water \cite{Huang08b}. However the experimental slip lengths show a  systematic bias toward higher values than theoretical ones. The discrepancy is actually not understood and might be the effect of gas adsorbed at the wall, possibly condensed in nanobubbles, an active reseach area recently \cite{LohseBorkentPRL07,LohseYangLA07}.

Finally, 
the case of organic molecules still raises debates. In the case of perfect wetting, various works on mica surfaces \cite{Chan85,Becker03}, as well as on glass surfaces \cite{Georges93,SteinbergerLA08},  showed that alkanes and OMCTS have a no-slip velocity either at the wall or at a molecular size inside the liquid phase (negative slip lengh). However Pit \etalns, using  a fluorescence recovery method with molecular probes in a TIRF configuration,
found very large slip lengths ($\simeq$ 100nm)  of hexadecane and squalane  on atomically smooth saphir surfaces although these liquids wet perfectly the saphir \cite{PitPRL2000, SchmadtkoPRL05}. Further work is under progress to understand if these large slip lengths in the case of favorable solid-liquid interactions reflect a particular effect of molecular structure and incommensurability at the saphir/liquid interface. 
More generally, although the problem of the boundary condition  has generated an imposant instrumental progress in the quantitative measurement and control of  flows at a nanoscale, a full quantitative  confrontation with the theory, \eg Eq.(\ref{formuleb}), is still lacking.


\section{Nanometric diffuse layers and transport therein}
\label{sec:DL}

The transport of electrolytes within the EDL and the associated phenomena have been extensively studied during the last century due in particular to its central 
role in colloid science. 
It has been accordingly discussed exhaustively in textbooks and reviews \cite{Hunter,Anderson89} and in the present section we merely point to
the main concepts which will be useful for our discussion. 

In spite of this long history, new directions and perspectives have been opened very recently in this old domain. 
Novel insight into the nanometric Debye layer have indeed been reached: new tools now make it possible to investigate the detailed structure and
dynamics {\it inside} the Debye layer. This concerns both the experimental side, see previous section, and numerical side, with the development
of ever more powerful Molecular Dynamics tools.

From these more 'molecular' views of the EDL, in contrast to the more standard 'continuum' picture of the EDL, new concepts and phenomena have emerged.
We shall thus merely focus here on the recent progress and new leads opened by nanofluidics in this domain. 



	\subsection{From Debye layer to ion specific effects}

An EDL builds up at a charged interface: it is the region of finite width where the surface charge is balanced by a diffuse cloud of counterions \cite{Anderson89}.
The Debye layer is usually described on the basis of the mean-field Poisson-Boltzmann (PB) theory, see \cite{Andelman95} for a review.
Ions are described as point charges and only their valency enters the description. Correlations between charges are accordingly neglected. 
The thermodynamic equilibrium balancing the ions entropy to their electrostatic interaction with the surface (attraction for the counter-ions and repulsion for the co-ions)
leads to the introduction of the Debye length, as introduced above in Eq. (\ref{Dlength}) and which we recall here:
\begin{equation}
\lambda_D= \left(8\pi \ell_B \rho_s\right)^{-1/2}
\end{equation}
Getting more into details, the PB description results from the combination of (i) the Poisson equation relating the electrostatic potential $V$ 
to the charge density $\rho_e=e(\rho_+-\rho_-)$ according to
\begin{equation}
\Delta V = - {\rho_e \over \epsilon}
\label{Poisson}
\end{equation}
with $\epsilon$ the dielectric constant, here identified to its bulk value, and $\Delta$ the spatial Laplacian; and (ii)
the thermodynamic equilibrium leading to a spatially constant electro-chemical potential for the ions:
\begin{equation}
\mu_{\pm}= \mu(\rho_\pm) \pm z_\pm e V
\label{EC}
\end{equation}
with $z_\pm$ the ion valency, $\rho\pm$ the ion densities and under the above assumptions, $\mu(\rho)=\mu_0+k_BT \log \rho$ the perfect gas expression for the 
chemical potential.

Though its crude underlying assumptions, the mean-field PB theory captures most of the physics associated with the EDL and is the usual
 basis to interprete ion transport.
One reason for this success is that ion correlations within the EDL, which are discarded in PB, can generally be neglected.  
This can be quantified by introducing a coupling parameter, $\Gamma_{cc}$, which compares the typical inter-ions electrostatic interaction to thermal energy
\cite{Levin03,Rouzina96}.
Following Refs \cite{Levin03}Êand \cite{Rouzina96}, the latter is estimated as $\Gamma_{cc} \approx e^2/(4\pi\epsilon \ell ) /{k_BT}=\ell_B/\ell$, with $\ell$ the mean distance between ions at the surface. This scale is
related to the charge density $\Sigma$, as $\ellÊ\approx \Sigma^{-1/2}$, so that $\Gamma_{cc} \approx \sqrt{\Sigma \ell_B^2}$ \cite{Rouzina96}. PB is expected to break down 
when the coupling parameter exceeds unity. For typical surfaces this threshold is not reached: for glass, the surface charge is at most in the $10^{-2}$C/m$^2$ range ($\sim 6.10^{-2}\, e/$nm$^2$),
and the coupling parameter $\Gamma_{cc}$ remains smaller than unity.

Other assumptions inherent to PB could be also questioned. In the PB description, the dielectric constant is assumed to be spatially uniform and identical to its bulk value. This
assumption is expected to break down very close to surfaces, in particular hydrophobic, where water ordering induce a local electric dipole pointing outward to the interface
\cite{Huang08}. This casts some doubts on the local relation between medium polarization and local electric field. However MD simulations suggest that
using the bulk dielectric constant in Poisson equation is not a critical assumption and can be safely used to describe the EDL \cite{Huang08}.

\begin{figure}[tb]
\begin{center}
\includegraphics [width=8 cm] {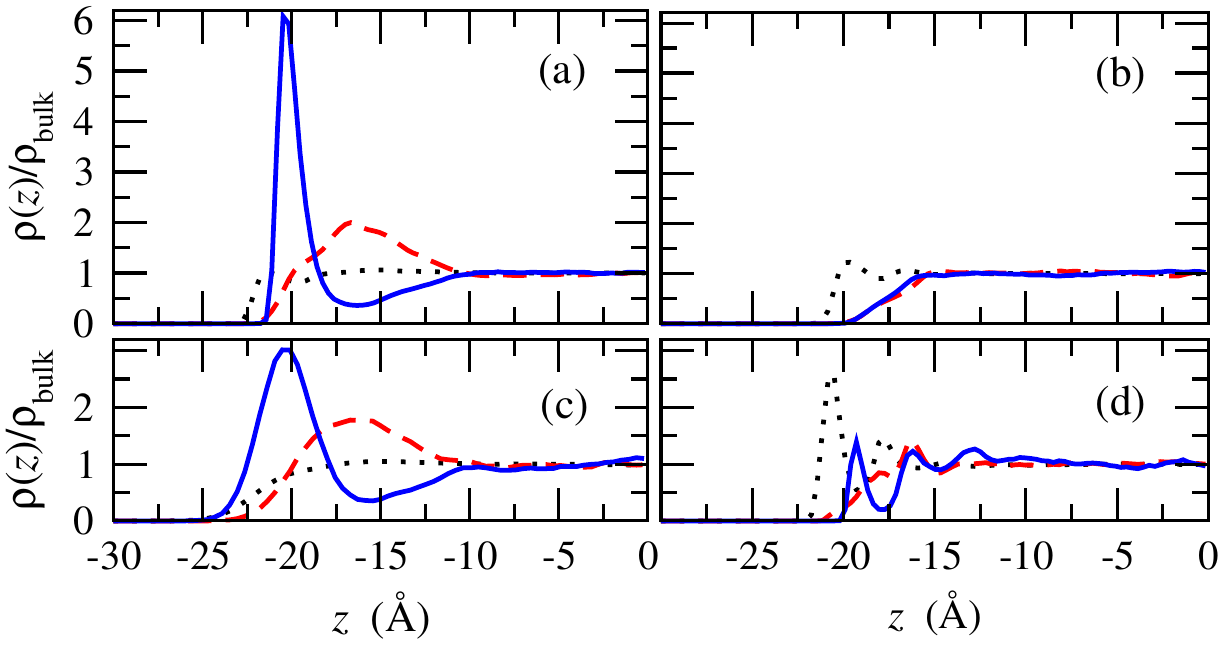}
\end{center}
\caption{Simulated density profiles of
negative ions (solid lines), positive ions (dashed lines), and water (dotted lines) 
 for roughly 1-M solutions of (a) NaI and (b) NaCl between
neutral hydrophobic surfaces, (c) NaI at a vapor--liquid interface, and (d) 
NaI between neutral hydrophilic surfaces.  From Ref. \cite{Huang07}.}
\label{fig:ionspec}
\end{figure}

PB description however misses an important class of effects associated with the specific nature of ions, and which affect the fine structure of the EDL.
Ion specificity is intimately connected to Hoffmeister effects, namely that  the
interactions between charged and neutral objects in aqueous media do depend crucially on the type of ion and not only on
electrolyte concentration \cite{Kunz04,Horinek07,Jungwirth06}. Such effects occurs not only at the air-water interface, where it does affect surface tension depending on ion type,
but also at hydrophobic surfaces \cite{Huang07,Horinek07}. 
This effect is evidenced in Fig.~\ref{fig:ionspec}, where results of molecular dynamics simulations of ions dispersed in water (SPC-E model) from Ref. \cite{Huang07} 
demonstrate that iodide ions do strongly adsorb at air-water and hydrophobic interfaces, while chloride do not. Note that a EDL builds up even in the case of a neutral
interface, we shall come back on this result and its implications.
In general heavier halide ions adsorb at the hydrophobic surface. This effect is mostly absent at a hydrophilic interface.
Tools of molecular simulation and new spectroscopic techniques have fully renewed the interest in ion specific phenomena \cite{Jungwirth06}.
The origin of the effect involves many aspects, as ion polarizability, image-charge interaction, ion-size (steric) effects, dispersion forces and the mechanisms are
still the object of intense research ... and also of some fierce debate, such as for the prefered adsorption of hydronium versus hydroxyl at interfacesÊ\cite{Buch07}, 
which fixes the charge of the liquid-vapor interface.

Such ion-specificity effects, which go beyond the traditional PB framework, have profound effects on the ion-transport process within the Debye Layer,
which we explore in the next part.
	
	\subsection{Interfacial transport: Electro-osmosis , streaming currents, and beyond}
	\label{sec:EOslip}
The EDL is at the origin of various electrokinetic phenomena. Here we shall discuss more particularly electro-osmosis (and -phoresis for colloids) and
its symetric phenomena, streaming currents. Both phenomena take their origin in the ion dynamics within the EDL. Due to the nanometric
width of the EDL, these phenomena are '{\it nanofluidic}' by construction and the various lengths introduced in Sec.\ref{sec:lengths} will appear.

Let us first discuss electro-osmosis (EO). EO is the phenomenon by which liquid flow is induced by an electric field.
In its simplest form, the fluid velocity $v_f$ beyond the EDL is linearly connected to the applied electric field $E_e$ according to the Smoluchowki
formula \cite{Hunter}:
\begin{equation}
v_\infty= - {\epsilon \zeta \over \eta} E_e
\label{Smolu}
\end{equation}
with $\epsilon$ the dielectric constant, $\eta$ the viscosity and $\zeta$ the so-called zeta (electrostatic) potential. The common interpretation of
$\zeta$ is that it is the electrostatic potential at the ``shear plane'', \ie the position close to the wall where the hydrodynamic velocity vanishes.
As we now discuss, this interpretation should be taken with caution.

Let us briefly recall the derivation of Eq.\ref{Smolu}. Fluid dynamics is described by the stationary Stokes equation:
\begin{equation}
\eta {\partial^2 v_x \over {\partial z^2}} + \rho_e E_e =0
\label{Stokes}
\end{equation}
Here we use a frame where the electric field $E_e$ is along $x$ and parallel to the planar surface, and the velocity field $v_x$ only depends on the
direction $z$ perpendicular to it.

A key remark, already mentionned above, is that the charge density, $\rho_e=e(\rho_+-\rho_-)$, is non vanishing only within the EDL, so that
the driving force for fluid motion, $F_e=\rho_e E_e$ is limited to that nanometric region, and vanishes otherwise.
In order to obtain the velocity profile, the Stokes equation should be integrated twice, and thus requires two boundary conditions.
Far from the surface, the velocity profile is plug-like and $\partial_z v\vert_{z=\infty}=0$. At the wall surface, the hydrodynamic boundary conditions
for the hydrodynamic velocity profile should be specified. We have discussed above that the latter introduces a slip length $b$, according to
Eq.~\ref{slip}, as $b\, \partial v\vert_{z=0}=v(z=0)$. Using Stokes and Poisson equations, 
one gets immediately that the full velocity profile is related to the electrostatic potential $V$ as
\begin{equation}
v(z)=- {\epsilon \over \eta} E \times [ -V(z) + \zeta  ]
\label{vel}
\end{equation}
where $\zeta$ has the meaning of a zeta potential\footnote{We choose here to define the zeta potential as the one appearing in the Smoluchowski
formula. According to Eq.\ref{zetab}, this may differ from the surface potential, depending on conditions. Another possibility would have been to define the zeta potential as
$V_0$ and consider an amplified zeta potential. We prefered to use the former definition.} and takes the expression
\begin{equation}
\zeta=V_0\times(1+b\cdot\kappa_{\rm eff})
\label{zetab}
\end{equation}
where $V_0$ is the electrostatic potential at the wall and $\kappa_{\rm eff}$ is the surface screening parameter, defined as 
$\kappa_{\rm eff}=-V^\prime(0)/V_0$ \cite{Joly04,Huang08}. For weak potentials this reduces to the inverse Debye length: $\kappa_{\rm eff}\simeq \lambda_D^{-1}$, 
but its full expression takes into account possible non-linear effects.
The fluid velocity far from the surface (\ie outside the EDL) has the Smoluchowski expression above, Eq. (\ref{Smolu}), with the expression of the zeta potential, $\zeta$, given
in Eq.~\ref{zetab}. This prediction was first discussed in Ref. \cite{Muller86}, and rederived independently in more recent works \cite{Joly04,Huang08}.
We emphasize that the above result, Eqs.~(\ref{vel}-\ref{zetab}), does not make any assumption on the specific model for the EDL, except for a constant
dielectric constant $\epsilon$ and the validity of Stokes equation, which was discussed above to be valid down the nanometer.

\begin{figure}[htb]
\begin{center}
\includegraphics [height=3 cm] {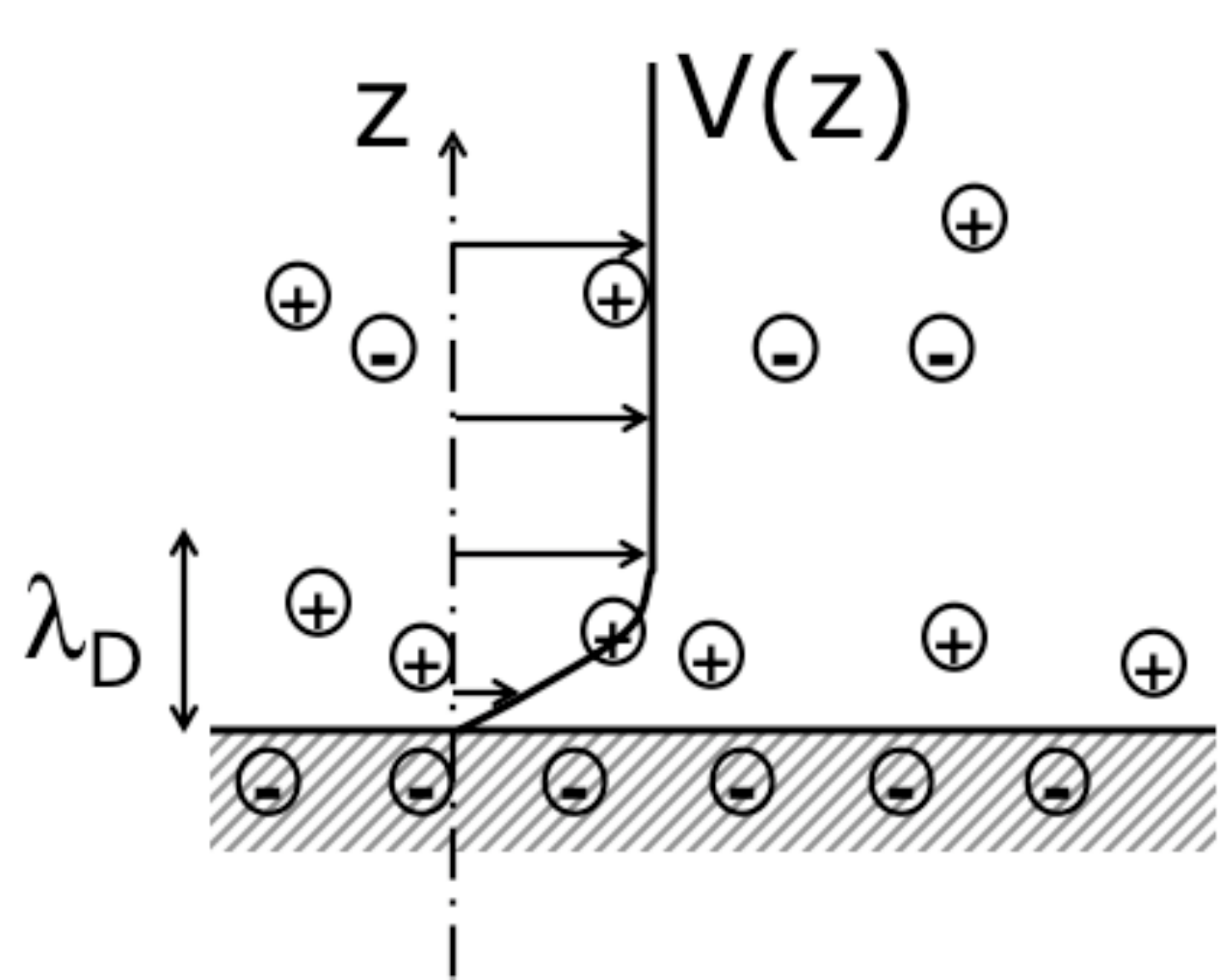}
\includegraphics [height=3 cm] {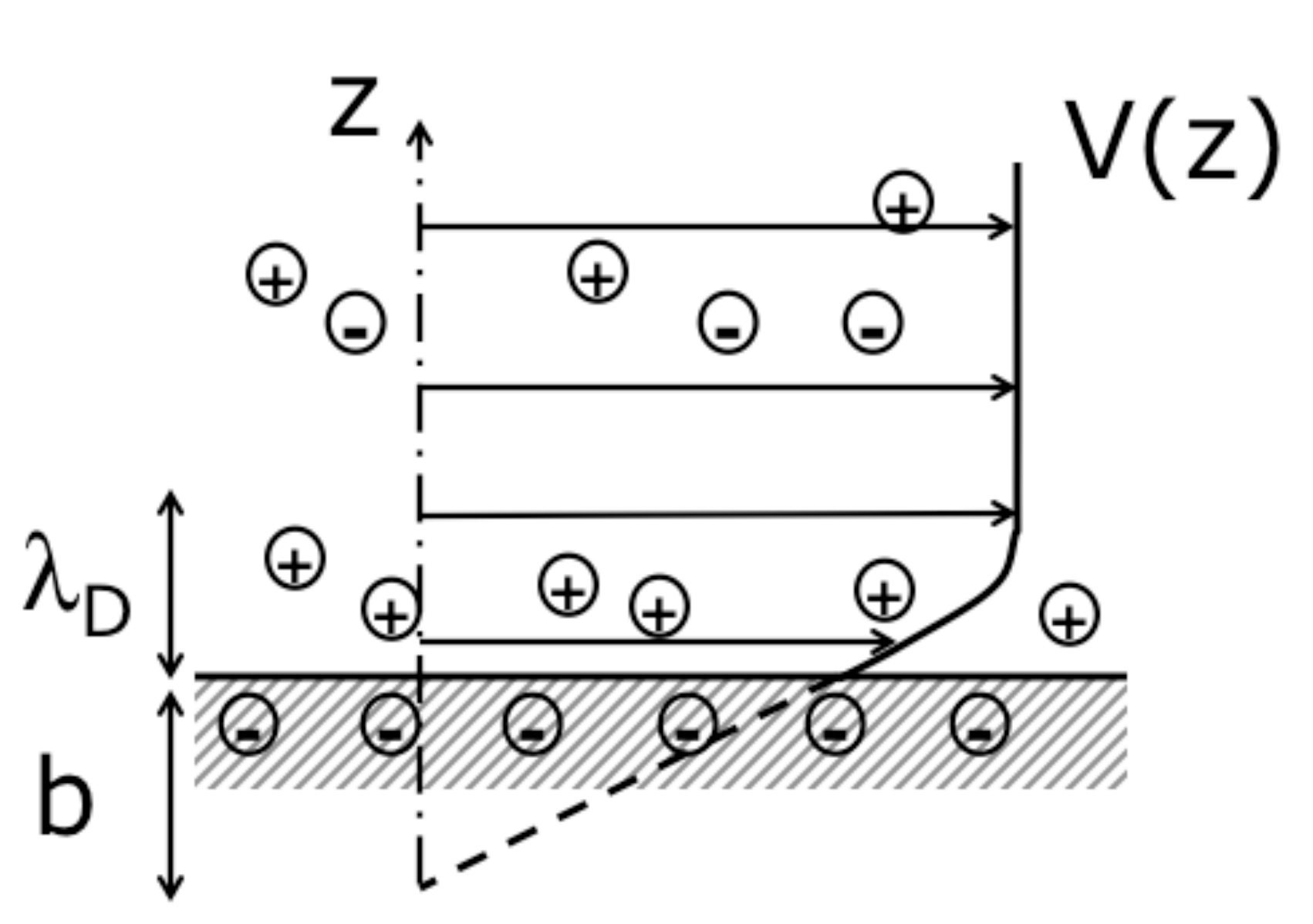}
\end{center}
\caption{Sketch of the influence of slippage on the electro-osmotic transport. Slippage reduces the viscous friction in the electric Debye layer, as the hydrodynamic velocity gradient occurs on a length
$b+\lambda_D$, instead of $\lambda$ without slippage. Flow is accordingly enhanced by a factor
$1+b/\lambda_D$. }
\label{fig:EOslip}
\end{figure}

The physical meaning of the above expressions is quite clear. The velocity in the fluid results from a balance between the driving electric
force and the viscous friction force on the surface. Per unit surface, this balance writes 
\begin{equation}
\eta {v_\infty\over \lambda_D + b} \approx e\Sigma \times E
\end{equation}
with $\Sigma$ the surface charge density (in units of $m^{-2}$) at the wall. The expression for the viscous friction expresses the fact that flow gradients occur on a length $b+\lambda_D$.
Relating the surface charge to the surface potential $V_0$ ($e \Sigma=-\epsilon\, \partial_z V\vert_0$), one recovers the above results.

As a side remark, one may note that the slip enhancement does not require any assumption on the electric and fluid behavior.
Indeed at a fully general level, the (exact) momentum balance at the surface  shows that the velocity at the wall is {\it exactly} given by $v_{\rm wall}= {D_{\rm wall}\over \lambda}ÊE$, with $D_{\rm wall}$ the dielectric displacement
computed at the surface, and $\lambda$ the surface friction coefficient introduced in Eq.(\ref{friction}). The latter is related to slip length according to the definition $\lambda\equiv b/\eta$. This expression for the wall velocity under an applied electric field neither invokes Navier-Stokes equation, nor any assumption on the dielectric behavior. This wall velocity acts as a lower bound to the asymptotic EO velocity, \ie the one measured at infinite distance from the surface. This demonstrates that the enhancement of the EO velocity due to low liquid-solid friction, \ie large slippage, has a fully general validity.

Several conclusions emerge from the above equation. First if the wall is characterized by a no-slip BC, with $b=0$, then the zeta potential
$\zeta$, Eq. (\ref{zetab}) identifies with the electrostatic potential $V_0$ computed at the surface (up to a possible molecular shift due to the precise position of the hydrodynamic BC
which defines the so-called shear plane
\cite{Joly04}): this is the common assumption, as usually expressed in textbooks \cite{Hunter}. 
Now, if finite slippage occurs at the surface, then {\it the zeta potential is much larger than the surface electrostatic potential}, $V_0$: it is amplified by a factor $1 + b/\lambda_D$ 
(assuming for the discussion that $\kappa_{\rm eff}=\lambda_D^{-1}$). This factor may be {\it very large}, since on bare hydrophobic surfaces 
$b$ may reach a few ten nanometers (say $b\sim 20-30$nm), while the Debye lentgh typically ranges between 30 nm down to 0.3 nm (for $10^{-4}$M to 1M).

These predictions have been confirmed by molecular Dynamics simulations of electro-osmosis (and streaming current, see below) \cite{Joly04,Huang07,Huang08}.
Various systems were considered, from a model electrolyte involving ions in a Lennard-Jones solvent, to a more sophisticated SPC-E model of water, with 
ions of various nature. Altogether simulations confirm the key role of slippage on electrokinetics and results do fully confirm the above description.

\begin{figure}[h]
\begin{center}
\includegraphics [width=6 cm] {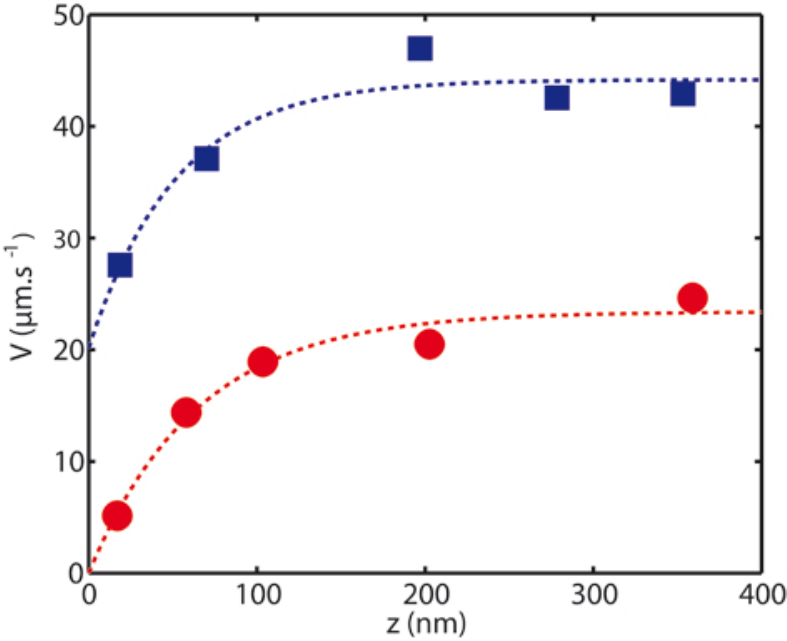}
\end{center}
\caption{NanoPIV measurements for the electroosmotic flows close to hydrophilic pyrex $\bullet$ and hydrophobic OTS-coated surface $\blacksquare$.
The deduced zeta potentials  are $\zeta=-66\pm 8$mV (hydrophilic glass) and $\zeta=-123 \pm 15 $mV (hydrophobic, silanized surface), while the electrostatic potentials were independently measured to be
comparable: $V_0\simeq-69$mV for the hydrophilic (pyrex) surface, and $V_0\simeq-65$mV for the hydrophobic (OTS-coated) surface.
The salt concentration
is $1.5\, 10^{-4}$mol.L$^{-1}$ and the driving electric field
is $E=500$V.m$^{-1}$. Dashed lines are fits to the theoretical predictions using Eq. (\ref{vel}), with  $\lambda_D=51\pm ${10} nm (measured independently) and $b=0\pm 10$nm (bottom) and $b=38\pm 6$nm (top) for the slip lengths. Figure from Ref. \cite{Bouzigues08}.}
\label{nanoPIV}
\end{figure}

On the experimental side, this strong influence of surface dynamics on the electrokinetics at surfaces has not been appreciated 
and explored in the electrokinetic litterature up to now, and 
very few experimental investigations have been performed on this question \cite{Churaev02,Bouzigues08}. One key difficulty is that the above result 
for the zeta potential, Eq. (\ref{zetab}), involves a 
strong entanglement between electrostatics, through surface potential, and fluid dynamics, through hydrodynamic slippage. In order to disentangle the two effects, 
two independent measurements of these quantities should in principle be performed. To our knowledge the first work on the subject was performed by
Churaev \etal in Ref. \cite{Churaev02}, and results indeed suggested a slip effect, at the expense however of a rather uncontrolled assumption on the surface potential.
More recently this problem was tackled in Ref. \cite{Bouzigues08} using the nanoPIV tool discussed above. This set-up allows for two independent measurements  of $\zeta$ and $V_0$, leading to 
an unambiguous confirmation of the above predictions of slippage effect on the zeta potential. 
Indeed  both the velocity profile, which 
is fitted to Eq.~\ref{vel}, and the nanocolloid concentration profile, from which the surface potential is deduced on the basis of the electrostatic repulsion, are measured independently.
Velocity profiles from Ref.\cite{Bouzigues08} are displayed in Fig.~\ref{nanoPIV}. While the two surfaces under consideration (a hydrophilic glass and a silanized hydrophobic surface) 
had basically the same surface potential, a factor of two is found on the zeta potential under the conditions of the experiment. Results are consistent with a slip length of $\approx 40$nm.
The factor of two for $\zeta/V_0$ occurs here due to a Debye length of $\approx 50$nm in the experimental condition of Ref. \cite{Bouzigues08} (a large Debye length
is required to investigate flow inside the Debye layer at the present spatial resolution, $\sim 20nm$, of the nanoPIV technique).
As a side remark, note the large slip velocity at the wall in the slippy case in Fig.\ref{nanoPIV}: this is expected since according to Eq.\ref{vel}, the velocity at the surface
is directly proportional to the slip length $b$, as $v(0)=- {\epsilon \over \eta} E \times V_0 {b\over \lambda_D}$ (assuming, to simplify, that $\kappa_{\rm eff}\approx \lambda_D^{-1}$).

This result opens new perspective to make use of surface physico-chemistry in order to optimize electric-induced transport and control flow by surface properties.
It was also argued that such slip-induced optimization would strongly enhance the efficiency of energy conversion devices based on electrokinetic effects \cite{Pennathur07,Ren08}.

Besides slip effect, ion-specificity has also an important influence on the electro-osmotic transport. As we reported above, the structure of the EDL
is affected by the nature of the ions under consideration, especially at hydrophobic walls (and air-water interfaces).
This can be best seen by rewriting the expression for the zeta potential, in Eq.\ref{zetab}, in a slightly different form:
\begin{equation}
\zeta= {1 \over \epsilon} \int_0^{\infty} (z^\prime+b)\rho_e(z^\prime)\, dz^\prime
\label{zetabis}
\end{equation}
with $\rho_e$ the charge density.
This expression \eg comes from the double integration of the Stokes law. As seen in Fig. \ref{fig:ionspec} ion specific effects have a strong influence on the
charge distribution, and do indeed strongly modify the value of $\zeta$.  This effect was studied in various MD simulations of electrokinetics
\cite{Qiao04,Huang07,Huang08} and several counter-intuitive effects were indeed observed, such as flow reversal (as compared to the expected surface charge) \cite{Qiao04},
or even the existence of a non-vanishing zeta potential for a {\it neutral} surface \cite{Huang07,Huang08}~! These effects can be understood from 
Eq.~\ref{zetabis}: while for a vanishing surface charge, the system is electro-neutral and $\int_0^{\infty} \rho_e(z^\prime)\, dz^\prime =0$, this however does not imply
that its first spatial moment, which enters the expression of $\zeta$ according to Eq.~\ref{zetabis}, be zero.

More quantitatively, effects of ion-specificity on the zeta potential were rationalized  in Ref. \cite{Huang08} on a basis of a minimal model, which was subsequently validated on MD simulations. 
This description is based on the idea that 'big' ions have a larger solvation energy and are therefore attracted to hydrophobic interfaces in order to minimize this cost \cite{Horinek09}.
An estimate of the solvation free-energy of ions close to interface allows to propose analytical predictions for ion-specific effects on the zeta potential. 
\vskip0.5cm

We finally quote that all the above discussion on the electro-osmotic transport also applies to the streaming current phenomenon, 
by which an electric current is induced by the application of a pressure gradient. Again, since the EDL is not electro-neutral, liquid flow will
induce an electric current inside the EDL \cite{Hunter}.
According to the Onsager symetry principle, these two phenomena -- electro-osmosis and streaming current -- are intimately related and the
cross coefficients should be identical.
The expression for the streaming current $I_e$ under an applied pressure gradient $\nabla P$ takes the form:
\begin{equation}
I_e = - {\epsilon \, \zeta \over \eta}Ê{\cal A} \left(-\nabla P\right)
\end{equation}
where $\zeta$ takes the same expression as in Eq.(\ref{zetab}) and ${\cal A}$ is the cross area of the channel.

We end up this section by mentioning that the above effects generalize to any interfacial transport phenomena. Indeed the above electrically
induced phenomena  belong to a more general class of surface induced transport, which also involves phenomena known as diffusio-osmosis or thermo-osmosis 
and their associated phoretic phenomena for colloidal transport \cite{Anderson89,Bocquet07}. Thermo-osmosis points to fluid motion induced by thermal
gradients, while diffusio-omosis corresponds to fluid flow induced by 
gradients in a solute concentration. Though relatively old, thermophoretic transport has been the object of recent investigations in particular for colloid manipulation \cite{Piazza08,Duhr06,Sano09}. Its origin is still the object of an intense debate with recent progress in disentangling the various contributions \cite{Duhr06,Piazza08,Wurger08}.
On the other hand, diffusiophoretic and osmotic transport have been less explored \cite{Anderson89,Ebel88} but their potential in the context of microfluidic applications
has been demonstrated recently \cite{Abecassis08,Huang08a}. This opens novel possibility of driving and pumping fluids \cite{Huang08a} -- as well as manipulating colloids -- with
solute contrasts, which have been barely explored up to now and would deserve further investigations in the context of nanofluidics. 
A common point to all these phenomena is their ``nanofluidic root'': as for the electro-osmotic transport discussed above, the driving force for fluid motion is
localized within a diffuse layer of nanometric size close to the surfaces. Therefore, this opens the possibility of strongly amplifying their effect on the basis
of slippage effects at the solid interface. As demonstrated theoretically in Ref. \cite{Ajdari06}, the amplification is -- as for electro-osmosis above --
amplified by a factor $1 + b/\lambda$, where $\lambda$ is the width of the diffuse layer. Similar effects are predicted for thermophoretic transport \cite{Morthomas09}.

Finally, one may raise the question of interfacial transport on {\it super-hydrophobic} surfaces. Such surfaces, achieved using
nano- or micro- engineering of the surfaces, were shown to considerably enhance the slippage effect and exhibit very large slip length in the micron (or even larger) range \cite{Joseph06,Ybert07,Lee08}. Accordingly, a naive application of the previous ideas would suggest massive amplification by a factor up to $10^4$ (!). However, the composite structure of the superhydrophobic interface, involving both solid-gas and liquid-gas interfaces, makes these transport mechanisms far more complex in this case than on smooth interface. It has been shown that in the regime of thin Debye layer, no amplification is obtained for electro-osmosis on superhydrophobic surfaces \cite{Huang08a,Squires08}, unless a non-vanishing charge exists at the
liquid-gas interface. In contrast, a massive amplification is predicted for diffusio-osmosis \cite{Huang08a}. This strong prediction
has not received an experimental confirmation up to now.


Furthermore, a massive amplification, by a factor up to $10^4$ (!), may be achieved when using nano-engineered surfaces associated with super-hydrophobic properties, which considerably enhance the slippage effect (with $b$ raising to the micrometers range) \cite{Joseph06,Ybert07,Lee08}.

		\subsection{Surface versus bulk: conductance effects}
	\label{sec:surfcond}	
	
We now discuss the surface conductance effect in nanochannels. 

The conductance, $K$, characterizes by definition the electric current versus electric potential drop relationship.  
As pointed out previously, conductance probes the number of free charge carriers, here ions.
In the bulk $K$ is therefore expected to be proportional to the salt concentration $\rho_s$ (ionic strength).
Now in the presence of surfaces, the charges brought by the surface lead to a supplementary contribution to the conductance. 

Surface conductance is actually a rather classical phenomenon in colloid science \cite{Hunter}, but it was
demonstrated quite recently in the context of nanochannel transport by Stein \etal \cite{Stein04} and then by other groups \cite{Karnik05b}.


As we introduced in Sec. \ref{sec:electrostat}.c, the ratio of bulk to surface charge carriers is characterized by the Dukhin length $\ell_{Du}$. This length accordingly describes the competition between bulk and surface conductance. 
Depending on salt concentration, this length can be much larger than the Debye length.
Surface conductance is therefore expected to dominate over the bulk contribution for nanochannels smaller than $\ell_{Du}$. 
The magnitude of the surface conductance effect is amplified in small channels, but a key point is that it does not require Debye layer overlap.
%

Experimentally, the surface contribution to the conductance shows up as a saturation of the conductance 
in the limit of small salt concentration, 
while its expected bulk counterpart is expected to vanish in this limit.
The saturation originates in the charge carriers brought by the surface charge, the number of which is independent of
the salt concentration. 
This is illustrated in Fig.~\ref{fig:Stein}, from \cite{Stein04}, where the conductance  is measured in channels of various width, from 1015nm down to 70nm:
\begin{figure}[htb]
\begin{center}
\includegraphics [width=7 cm] {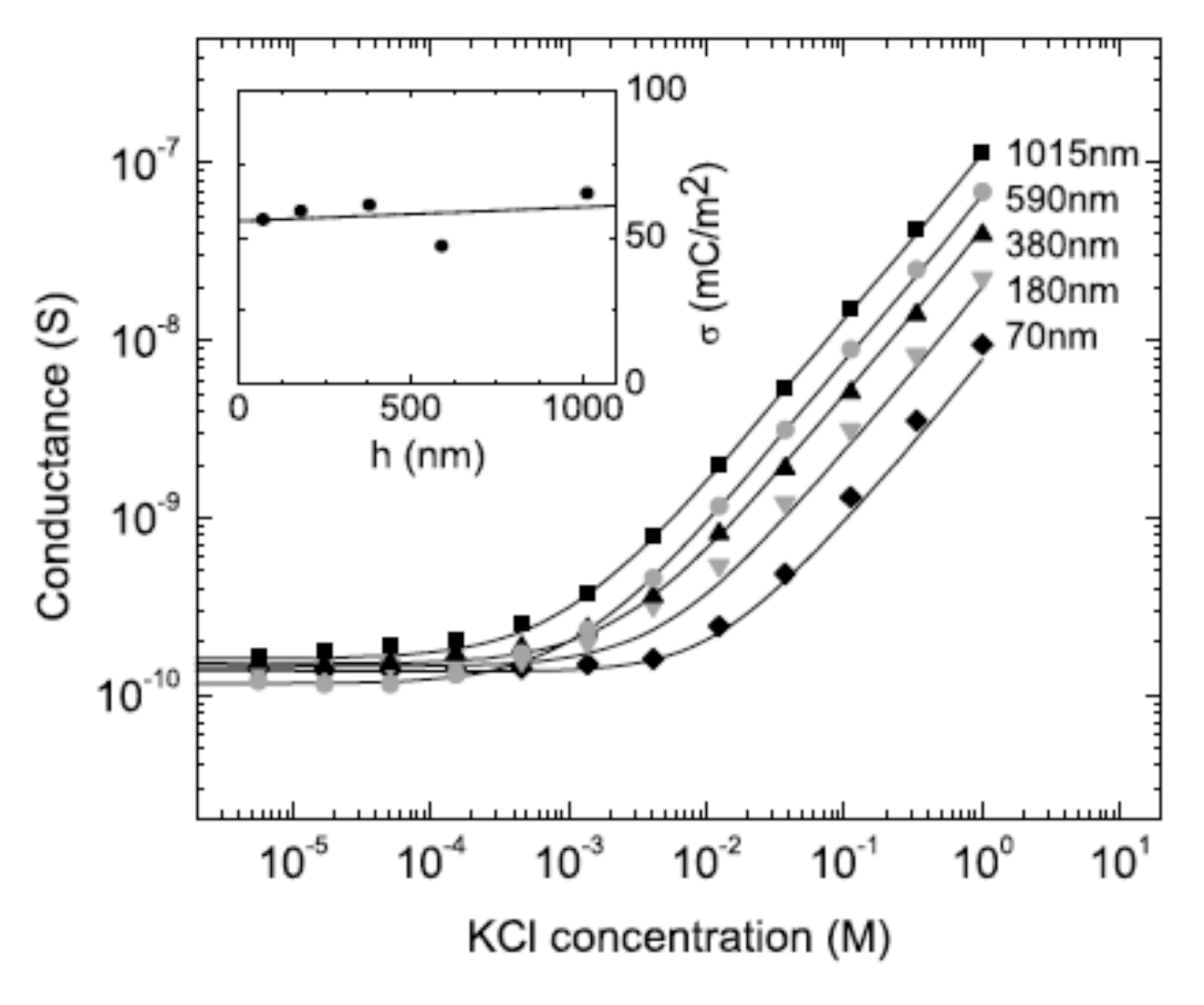}
\end{center}
\caption{Channel height dependence of ionic conductance
behavior. The conductance of fluidic channels is plotted against salt concentration for various nanochannel width $h$
($h=1015$, $590$, $380$, $180$, and $70$ nm). The inset displays
the fit values for the surface charge at the nanochannel walls as a function of $h$, from Ref. \cite{Stein04}.}
\label{fig:Stein}
\end{figure}

Let us discuss more specifically this point.
In a slit geometry with width $h$, the general expression for the current (for a unit depth of the slit) takes the form
\begin{equation}
I_e= e \int_{-h/2}^{h/2} dz\, \left[\rho_+ (z) u_+(z)-\rho_-(z) u_-(z)\right]
\label{Ie}
\end{equation}
where the averaged velocites of the ions is 
\begin{equation}
u_\pm(z)=v(z)\pm e \mu_\pm E_e,
\label{uu}
\end{equation} 
with $\mu_\pm$ is the ion mobility, defined here as the inverse of the ion friction coefficient (we assume furthermore $\mu_\pm=\mu$). 
The velocity $v(z)$ is the fluid velocity induced by electro-osmosis under the electric field, as given in Eq.(\ref{vel}).
The second term is the contribution to the current due to ion electrophoresis.

Combining the expression of the velocity $v(z)$, as in Eq.(\ref{vel}), to Eqs.~(\ref{Ie}-\ref{uu}), one may remark that the electro-osmotic contribution to the current
takes the form of the integral of  the charge density times the electrostatic potential: this has therefore
the form of an electro-static energy, ${\cal E}$. This is confirmed by the full calculation, which leads to a conductance expression in the form \cite{Stein04,Levine75}:
\begin{equation}
I/E_e=2 e^2\,\mu\, \rho_s  {\cal A} (1 + H)
\end{equation}
with ${\cal A}$ the channel cross section (${\cal A}=w\times h$ with $w$ the depth of the channel) and the surface correction $H$
\begin{equation}
H=\cosh\left({e V_{\rm c}\over k_BT}\right)-1 + {{\cal E} \over 2 \rho_sk_BT \, h}\left( 1+ {1\over 2\pi \ell_B\, \mu\, \eta} 
\right)
\label{H}
\end{equation}
where $V_{\rm c}= V(0)$ is
the electrostatic potential at the center of the channel, and ${\cal E}= {\epsilon\over 2} \int_{\rm slit} dz\, \left({dV\over dz} \right)^2$ the electrostatic energy (and not the {\it free} energy).  Note that writing $\mu=3\piÊ\eta d_{i}$ with $d_i$ the ion diameter, the last term in Eq.(\ref{H}) is typically of order $d_i/\ell_B$.
It originates in the surface-induced -- electro-osmotic  -- contribution to the conductance.

In the low salt concentration regime, $\rho_s\rightarrow 0$,
the length scales are organized in the order: 
\begin{equation}
\ell_{GC} < \lambda_D < \ell_{Du}, 
\end{equation}
since 
$\lambda_D/\ell_{GC}=\ell_{Du}/\lambda_D \gg 1$ for  $\rho_s\rightarrow 0$.
Saturation will occur as the channel width $h$ is smaller than the Dukhin length, $h<\ell_{Du}$, independently of the order of the $h$ and $\lambda_D$.
Debye layer overlap is accordingly {\it not} the source of the saturation of the conductance. In this regime, the second term in Eq.~(\ref{H}), associated
with surface contributions is dominant.

In this limit, the calculation of the electrostatic energy shows that ${\cal E} \sim \vert\Sigma\vert$. The fact that 
${\cal E}$ scales linearly in $\vert\Sigma\vert$
and not as $\vert\Sigma\vert^2$ -- as would be first guessed -- is due to the 
non-linear contributions to ${\cal E}$ in the Poisson-Boltzmann description. These become dominant in the $\rho_s\rightarrow 0$ limit where the Debye length
becomes larger than the Gouy-Chapman length. 
Altogether, this shows that for $\rho_s\rightarrow 0$, the conductance saturates at a value 
\begin{equation}
K_{\rm sat}Ê\approx  e^2\,\mu\, w\, \times 2\vert \Sigma\vert \left( 1+ {1\over 2\pi \ell_B\, \mu\, \eta} 
\right)
\label{Ksatapp}
\end{equation}
with $w$ the depth of the channel. This value is independent of both $h$ and $\rho_s$.
This saturation can be also partly understood in the context of Donnan equilibrium discussed below. In the limit $\rho_s\rightarrow 0$, only the counter-ion contributes
to the ion concentration in the channel, $\rho_++\rho_-\approx 2\Sigma/h$, see Eq.(\ref{Donnan}) below, so that the conductance reduces to $K_{sat}\approx e^2 \mu\,w\, 2\vert \Sigma\vert$.
This analysis however misses the electro-osmotic contribution to the conductance (second term in the brackets in Eq.~(\ref{Ksatapp})).

A few remarks are in order.
First, the dependence of the conductance on the surface properties, here the surface charge $\Sigma$, opens new strategies to tune the nanochannel transport properties, {\it via}Ê an external control. This has been used by Karnik \etal to develop a nanofluidic transistor \cite{Karnik05}, allowing to control the conductance of the nanochannel 
thanks to a gate voltage. 

Another interesting remark is that along the same line as in Sec.~\ref{sec:EOslip}, the electro-osmotic contribution to the conductance could be also amplified by slippage effects.  
This can be readily demonstrated by a direct integration of Eq.~(\ref{Ie}). Accordingly slippage effects add a new contribution to the conductance:
\begin{equation}
K_{\rm slip} = 2 e^2\, w\, {\Sigma^2 \over \eta} \times b
\end{equation}
Assuming Stokes law for ion inverse mobility ($\mu=(3\pi\eta d_{\rm ion})^{-1}$ with $d_{\rm ion}$ the ion size), then $K_{\rm slip}/K_{\rm sat}\sim b \times d_{\rm ion}\, \vert\Sigma\vert \approx b/\ell_{GC}$: since the Gouy-Chapman length $\ell_{GC}$ is typically nanometric (or less),  this enhancement is therefore very large, even for moderate slip length ($b\sim 30$nm) !

The channel resistance is accordingly decreased by the same factor in the low salt regime. This amplification of conductance by hydrodynamic slippage 
opens very interesting perspectives in the context of electrokinetic energy conversion, as was pointed out recently by Pennathur \etal \cite{Pennathur07} and Ren and Stein \cite{Ren08}. 
Moderate slippage, with slip length of a few ten nanometers, 
is predicted to increase the efficiency of the energy conversion up to 40\% (and of course even more with larger slip lengths).
This attracting result would deserve a thorough experimental confirmation.

	\subsection{Debye layer overlap and nanofluidic transport}
	\label{sec:DLnano}

The phenomenon of Debye layer overlap has already been widely explored in the nanofluidic litterature. As we pointed above, 
various reasons underlie this specific interest: (i) well controlled nanometric pores with size in the range of 20-100 nm can be produced using micro-lithography techniques \cite{Schoch08}:
this does indeed correspond precisely to the range of typical Debye lengths for usual salt concentrations (remember that 
$\lambda_D=30$nm for a salt concentration of $10^{-4}$M); (ii) the overlap of Debye length does indeed have a strong influence on ion transport,
so that novel transport effects emerge at this scale, with applications for chemical analysis.

The question of Debye layer overlap, and related phenomena, was discussed quite extensively in a recent review by Schoch \etal \cite{Schoch08}. Here
we thus 
focus  on the main guiding lines underlying this problem and point to the key physical phenomena at play.
In particular we shall discuss an illuminating analogy to transport in semi-conductors.
%

\subsubsection{Donnan equilibrium}

\begin{figure}[h]
\begin{center}
\includegraphics [height=4. cm] {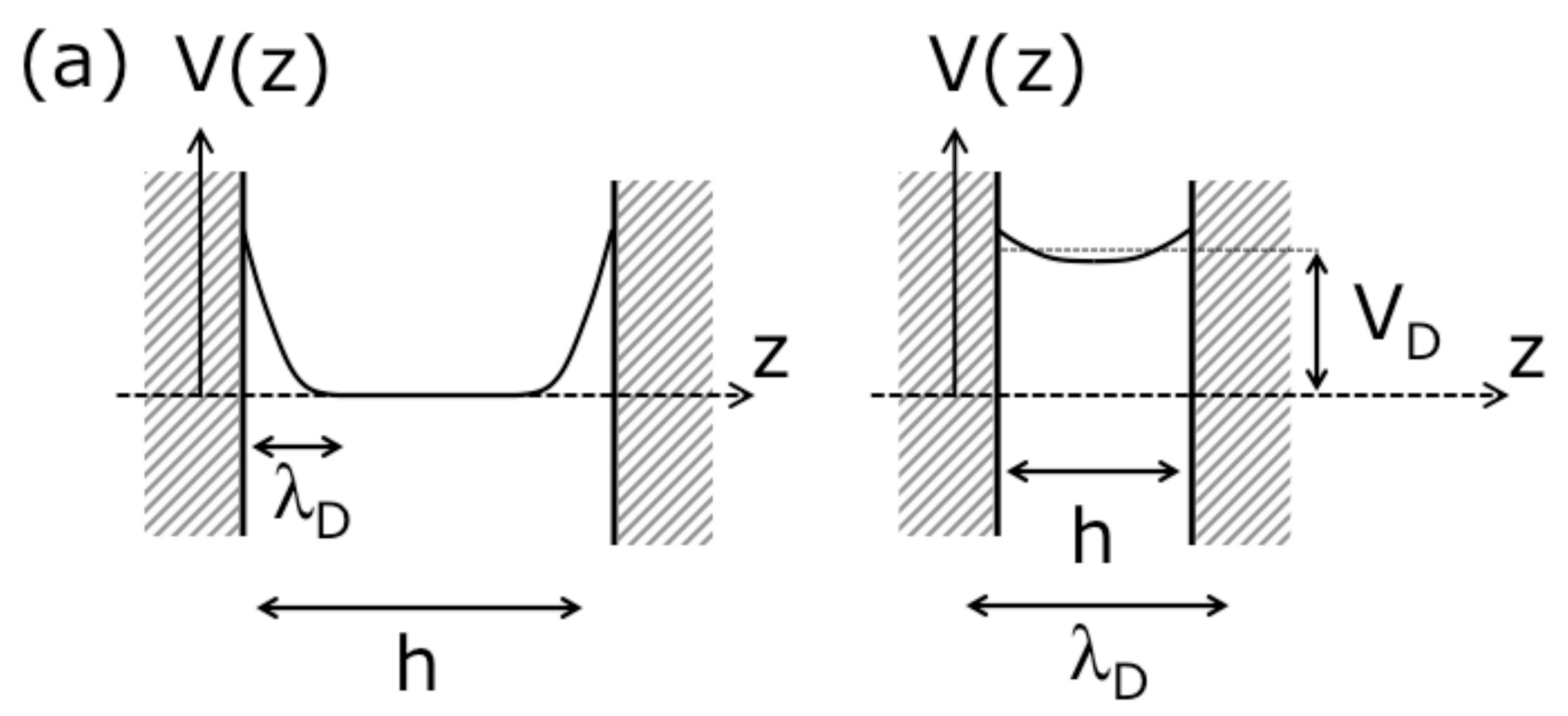}
\includegraphics [height=3.2 cm] {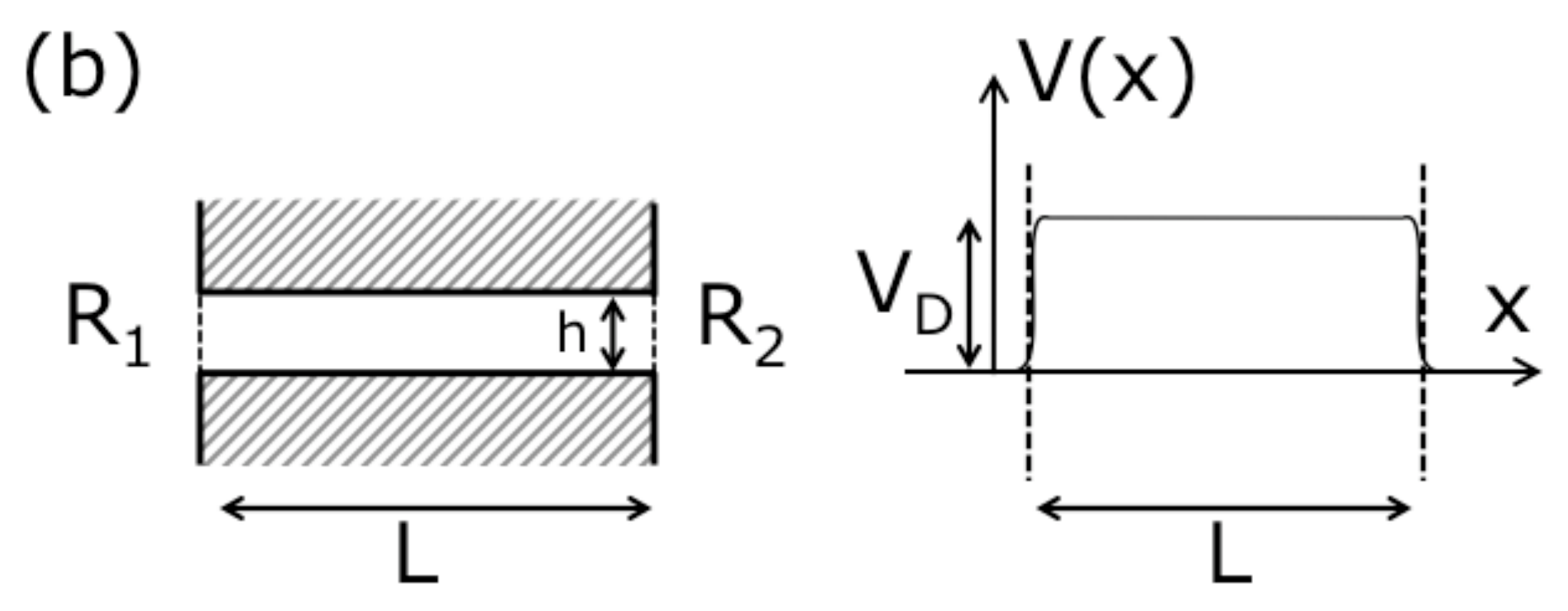}
\end{center}
\caption{(a): Sketch of the electrostatic potential in a charged slit with thickness $h$: ({\it left})  $h>2\lambda_D$; and ({\it right}), $h<2\lambda_D$, corresponding to Debye layer overlap. (b) A charged nanochannel, with thickness $h$ and length $L$, connects two reservoirs $R_1$ and $R_2$. In the situation of Debye layer overlap,  a Donnan potential $V_D$ builds up along the nanochannel.}
\label{fig:Donnan}
\end{figure}

 An important notion underlying the Debye overlap is the so-called ``Donnan equilibrium'', which is a well know concept
in the colloid litterature. Due to the supplementary charges brought by the nanochannel's surface, a potential drop builds up between
the nanochannel's interior and the external reservoir, in order to maintain a spatially uniform chemical potential of the ions.
The latter, as introduced above, takes the form (assuming a dilute ion system):
\begin{equation}
\mu(\rho_\pm)=\mu_0 + k_BT \log(\rho_\pm) \pm {e V} = \mu_0 + k_BT\log(\rho_s)
\end{equation}
with $\rho_s$ the (uniform) salt concentration outside the nanochannel, \ie the ``reservoir''.
These equations are
completed with the overall electroneutrality over the channel, 
\begin{equation}
\int_{-h/2}^{h/2}Êdz\, (\rho_+(z)-\rho_-(z))= 2\Sigma
\end{equation}
with $z$ along the direction perpendicular to the nanochannel (here assumed to be a slit),
see Fig. \ref{fig:Donnan}.

We now focus on the thin pore limit, assuming Debye layer overlap $h\lesssim2\lambda_D$. 
In this situation, ion densities and electrostatic potential
are approximately spatially uniform over the pore thickness $h$. 
Ion densities thus obey:
\begin{eqnarray}
&&\rho_+\times \rho_-=\rho_s^2 \nonumber \\
&&\rho_+-\rho_-={2\Sigma\over h}
\label{DonnanEqu}
\end{eqnarray}

Altogether, this leads to
\begin{eqnarray}
&&\rho_\pm=\sqrt{\rho_s^2+\left({\Sigma\over h} \right)^2}\pm{\Sigma\over h}\nonumber\\
&&V_D={k_BT\over 2e} \log\left[{\rho_-\over \rho_+}\right]
\label{Donnan}
\end{eqnarray}
where  the mean electrostatic potential $V_D$ is denoted as the Donnan potential.

Note that the amplitude of the Donnan potential $V_D$ is quantified by the ratio $\ell_{Du}/h$ where $\ell_{Du}=\Sigma/\rho_s$ 
is the Dukhin length introduced above. To fix ideas $\ell_{Du} \sim 1nm$ for $\rho_s=1M$ and a typical surface charge.


\vskip0.5cm
It is finally of interest to mention the strong analogy between the Donnan equilibrium
and the equilibrium of charge carriers in doped semi-conductors (SC). In the latter the 
electrons density $\rho_n$ and the holes density $\rho_p$ obey Eq. (\ref{DonnanEqu})
with  the electrolyte density $\rho_s$ replaced by the carriers density in the intrinsic 
(non-doped) SC.
The surface charge in the nano-channel actually acts as impurities  delivering additionnal
 carriers, with
$\Sigma >0$ corresponding to donnors (N-doped SC) and $\Sigma<0$ acceptors (P-doped SC).
The Donnan potential is then the analogous of the shift in the Fermi energy due to impurities, 
which determines the voltage difference at equilibrium between regions of different dopage level.
 The table hereafter, table I, summarizes the corresponding quantities in the nano-channel and SC analogy. 
 \begin{table}[!htbp]
   \centering

{
\scriptsize
{
\begin{tabular}{|c|c|}
    \hline
Nanofluidics &Semi-conductors\\
\hline \hline
$\rho_-$ & $\rho_n$\\
negative ions concentration & electron density in the conduction band\\
\hline
$\rho_+$ & $\rho_p$\\
    positive ions concentration  &holes density in the conduction band \\
   \hline
$\rho_s$ & $ n_i $\\
   electrolyte concentration & carriers density in the intrinsic SC\\
   \hline
surface charge $2\Sigma /h $& Impurities (dopping) concentration  \\
    \hline
   Donnan potential & Shift of the Fermi level \\
  \hline
\end{tabular}
}}\\
\caption{Equivalence table for corresponding quantities in doped semiconductors and nano-channels.}
\label{tab-SC}
\end{table}

\par

The Donnan potential and Debye layer overlap have many implications on nanofluidic transport. 

\subsubsection{Ion transport and PNP equations}

A classical framework to discuss ion transport in narrow pores is the so-called Poisson-Nernst-Planck (PNP) equations.
This simplified description of ion transport in strongly confined channels is based on the coupled diffusion-electro-convection of the ions \cite{Daiguji04}. 

As in Fig.\ref{fig:Donnan}, the channel thickness is assumed to be small compared to the Debye length, $h\ll \lambda_D$, so that 
ion concentrations are assumed to be uniform across its thickness. 
The ion fluxes $J_\pm$ (per unit surface) then takes the form
\begin{equation}
J_\pm=-D_\pm \partial_x \rho_\pm(x) \pm \mu_{\pm} e \, \rho_\pm(x) (-\partial_x V)(x)
\label{PNP}
\end{equation}
where $D_\pm$ and $\mu_\pm$ are related by the Einstein relationship, $D_\pm=\mu_\pm k_BT$;
$x$ is along the direction of the channel. Note that for simplicity, a single salt and monovalent ions are considered here. 
A local version of the electroneutrality, Eq. (\ref{DonnanEqu}), may be made along the pore:
\begin{equation}
\rho_+(x)-\rho_-(x)={2\Sigma(x)\over h}
\label{EN}
\end{equation}
Finally at the pore entrance and exit, a Donnan electric potential drop builds up, along the description given above, see Eq.~(\ref{Donnan}).

To illustrate further this approach, we consider the transport through a single nanochannel, as sketched in Fig.\ref{fig:Donnan}-b. At equilibrium the two reservoirs, $R_1$ and $R_2$ have the same electric potential and salt
concentration, associated with a uniform Donnan potential along the channel. Now if an
electrostatic potential drop $\Delta V$ or a salt concentration difference $\Delta\rho_s$ is imposed between the reservoirs,  ion fluxes will build-up so as to relax towards equilibrium.

In the stationary state, the ion flux are spatially uniform, so that one may solve Eqs. (\ref{PNP}-\ref{EN}) for the ion densities and electrostatic potential. 
This allows to compute
the current $I={\cal A}\,Êe (J_+-J_-)$ and total ion flux $\Phi_t={\cal A}\,Ê (J_++J_-)$, with ${\cal A}$ the cross area.
In general, the relationship $I(\Delta V,\Delta\rho_s)$, $\Phi_t(\Delta V,\Delta\rho_s)$ are non-linear.

However, in the limit of small  $\Delta V$ and $\Delta\rho_s$, a linear relationship can be written in general between the ionic
fluxes and corresponding thermodynamic 'forces' \cite{Groot69,Brunet04}:
\begin{equation}
\left[\begin{array}{c}I \\ \Phi_t\end{array}\right]
= {{\cal A}\over L}
\left[\begin{array}{cc}K & \mu_{K} \\\mu_{K} & \mu_{\rm eff} \end{array}\right]
\left[\begin{array}{c}-{ \Delta V} \\-{k_BT}\,\Delta[  { \log \rho_s]}\end{array}\right]
\label{Mat-transport}
\end{equation}
with ${\cal A}$ the cross area of the channel and $\Delta[ \log \rho_s]=\Delta\rho_s/\rho_s$.
Note that due to Onsager (time reversal) symetry, the non diagonal coefficients of the matrix are equal \cite{Groot69,Brunet04}.

The coefficients of the above matrix can be calculated within the PNP framework. PNP equations are solved analytically for  small potential and concentration drops, leading to the following expression for the various coefficients in the matrix, Eq.(\ref{Mat-transport}):
\begin{eqnarray}
&&K=2 \mu e^2\, \sqrt{\rho_s^2+ \left({\Sigma\over h}\right)^2}\nonumber \\
&&\mu_{\rm eff}=2 \mu \times \sqrt{\rho_s^2+ \left({\Sigma\over h}\right)^2}\nonumber \\
&&\mu_K=e\,\mu  \times {2\SigmaÊ\over h}
\label{mobility}
\end{eqnarray}
We assumed here that ions have the same mobilites $\mu_\pm=\mu$. The cross effects, associated with the mobility
$\mu_K$,  originate in the dependence of the Donnan potential on the salt concentration.
\vskip1cm

Let us conclude this part with a few remarks:\\ \\
\indent$\bullet$
First one should realize
that a number of assumptions are implicitly made in writing PNP equations. In particular the ions are treated as a perfect gas and  correlations between ions along the channel are neglected. This may become problematic in strongly confined situations
where single file transport (of the solvent) and strong unidirectional electrostatic correlations should build up. 
As an example, proton transport in single file water has been shown to involve a highly-cooperative mechanism \cite{Dellago03}.
However these limitations are restricted to single file transport in molecular channels, and should not be a limitation for pore size larger than a nanometer.
Furthermore the PNP model is interesting to explore as guiding line, in the sense that it provides a rather correct physical idea of the (complex) electro-diffusion couplings.
\vskip0.5cm
\indent$\bullet$
 Furthermore, we may pursue  the analogy discussed above between nanochannels and doped SC (see Table I),  and extend
 the discussion to transport phenomena. Indeed  {\bf the Poisson-Nernst-Planck equations, Eqs.(\ref{PNP}), are formally identical to the phenomenological transport equations for electrons and holes in SC}. Thus, as long as  electrical and concentration fluxes {\it only}
 are allowed, {\it nanofluidic devices can in principle reproduce standard SC-based components}, such as diodes and transistors. The nanofluidic diode for instance is based on the properties of a PN junction, \ie the junction between two regions with different doping: according to the equivalence table, Tabel I,  its nanofluidic equivalent corresponds to two nanochannels with different surface charge $\Sigma/h$. We will discuss below (paragraph 5) some recent findings confirming the pertinence of this analogy. 
 \vskip0.5cm

 \indent$\bullet$
 However one should keep in mind that the transport analogy between fluids and electrons breaks down in the presence of an hydrodynamic  flow. Actually,
 PNP equations introduced above do not take into acccount convective contributions. In a fluidic system, a flow is indeed expected to occur in nanochannels as soon as a voltage drop is applied to its end, due to the body force acting on the mobile charges, but also under a pressure gradient (hydrostatic or osmotic). 
 This is for example the source of electro-osmotic or streaming current contributions discussed above. In contrast, SCs can not flow - quite obviously - under an applied electric field and may only deform elastically.
 This is therefore a limitation to the above analogy.  
 
Beyond this analogy to SC transport, the above PNP equations do not involve any hydrodynamic contributions. This concerns for example the electro-osmotic contributions
(see \eg the very last -- viscosity dependent -- term in the expression of the conductance, Eq.(\ref{H})), or  
the slip-induced enhancement of these phenomena. These may well become dominant in the strongly confined regime for sufficiently large slippage.
In general a convection contribution to the ion flux should therefore be added to the above ion flux, in the form
$J_{\rm conv}^\pm=v_f \times \rho_\pm$, where $v_f$ is the water velocity. The water velocity is in turn coupled
to (electric) body forces via the momentum conservation equations,  in practice the Navier-Stokes equation in its
domain of validity.
Altogether, 
the hydrodynamically induced cross phenomena, \ie  electro-osmosis, streaming currents,  diffusio-osmosis, ...  (as well as bare 
osmosis),  can be summarized at the linear level in terms of a symetric transport
matrix, in a form similar to Eq.~(\ref{Mat-transport}). This matrix  relates linearly the fluxes to thermodynamic forces at work \cite{Kedem61,Kedem63,Brunet04}.
These phenomena have been discussed extensively in the context of membrane transport but generalize here to transport in nanofluidic channels. 

More generally the coupling of the ion transport to the fluid transport thus opens the route to new applications  in nanofluidic devices, with a richer phenomenology than in SC electronics.



\vskip0.5cm
We now illustrate the above concepts on a few practical situations. 

\subsubsection{Permselectivity}

\begin{figure}[htb]
\begin{center}
\includegraphics [width=5 cm] {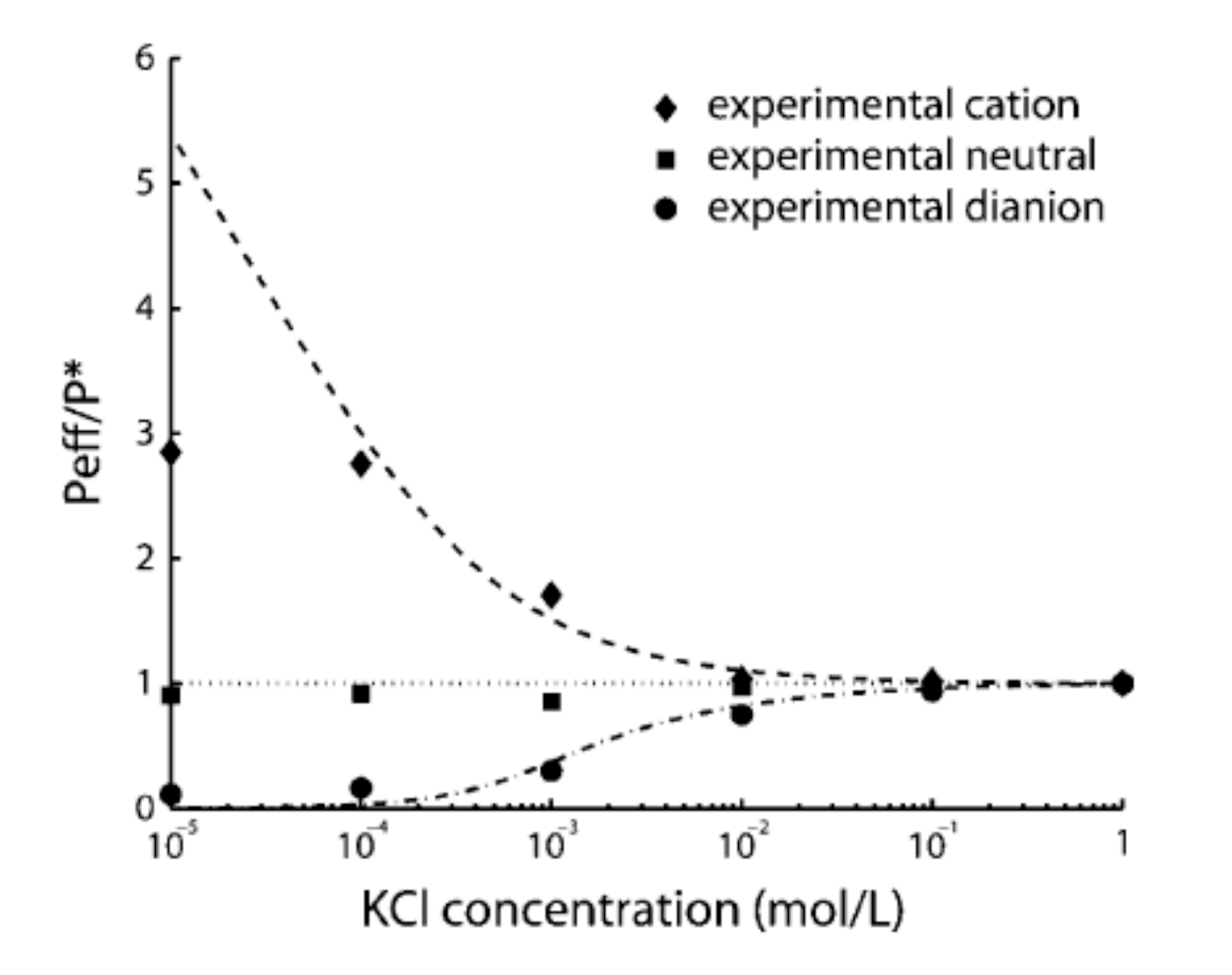}
\includegraphics [width=5 cm] {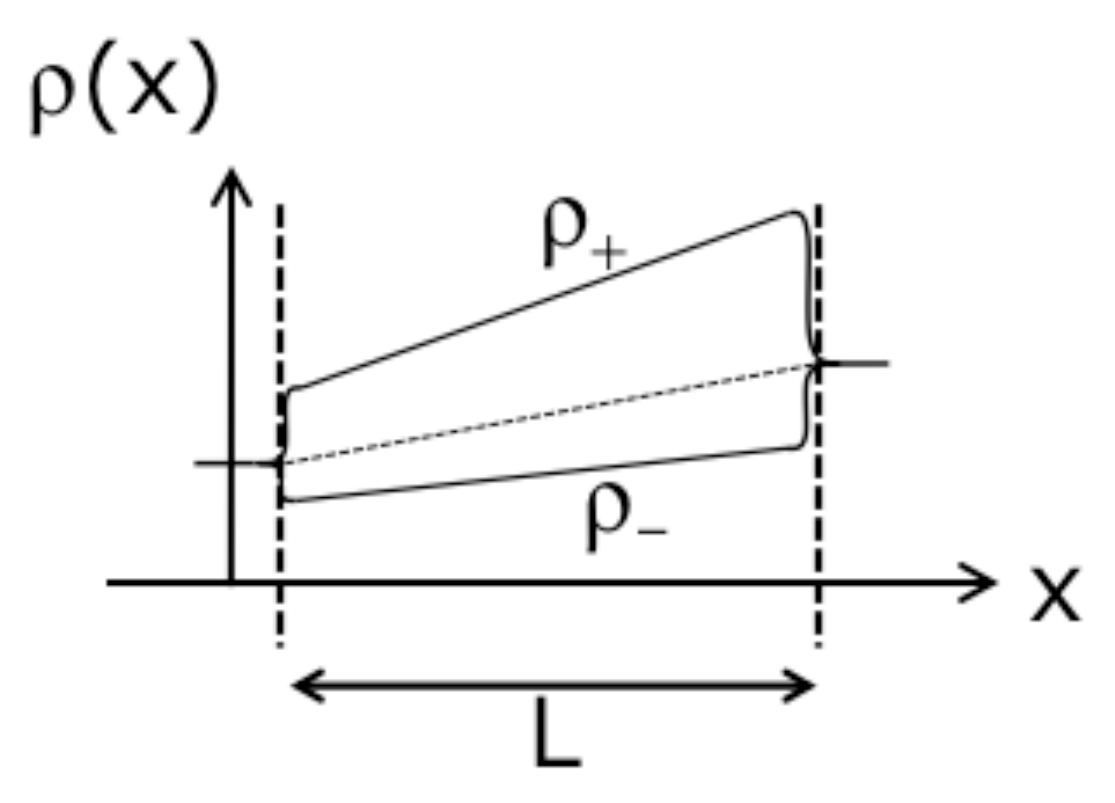}
\end{center}
\caption{  {\it Top}  Variation of the relative permeability of various probes with 
different charges, versus ionic strength. 
The experimental results are compared to 
theoretical fittings obtained with Eq.~(\ref{plecis}) (dotted lines). 
 {\it Bottom}Ê Concentration profiles of counter- and co-ions, $\rho_\pm$, for a given salt concentration drop
 $\Delta \rho$ between the two ends of the
nanochannel. The dotted line represents the linear concentration profile
for noninteracting diffusing species, whereas plain lines show the
ion profiles in the nanochannel. 
Adapted from Ref. \cite{Plecis05}.}
\label{fig:Plecis}
\end{figure}

 As a first example, we consider the permeability of nanopore to ions. It was first shown by Plecis \etalÊ\cite{Plecis05}
 that nanochannels exhibit a selective permeability for ion diffusive transport, see Fig.~\ref{fig:Plecis}(a): 
ions of the same charge as the nanochannel surface (co-ions) exhibit a lower permeability, while ions of the opposite charge
(counter-ions) have a higher permeability through the nanochannel.

This charge specific transport is a direct consequence of a  non-vanishing Donnan potential in the nanochannel (and
Debye layer overlap). As pointed out above, Fig.~\ref{fig:Donnan}, counter ions exhibit a higher concentration in the nanochannel
as compared to that of a neutral specie, $\rho_+>\rho_{\rm n}$, while co-ions have correspondingly a lower concentration, $\rho_-<\rho_{\rm n}$.

The diffusive flux of counter-ion will be accordingly larger than its expectation for neutral species, and vice-versa for co-ions.
This leads therefore to a charge-specific effective diffusion coefficient $D_{\rm eff}$ for the co- and counter- ions.


Following Plecis \etal \cite{Plecis05}, the effective diffusion coefficient is defined according to the identity
\begin{equation}
J_\pm= - D {\Delta \rho_{\pm}\over L} \equiv - D_{\rm eff} {\Delta \rho\over L}
\end{equation}
taking into account the fact that the local concentration in the nanochannel differs from the one imposed at the two
ends of the reservoir.
The linearized Poisson-Boltzmann equation is then used to calculate  the ion concentration in the slit, $\rho_{\pm}$, as:
\begin{eqnarray}
&\beta_\pm=& {\bar \rho_{\pm}(x)\over \rho(x)}= {1\over h} \int_{\rm slit} dz\, \exp[\mp eV(z)/k_BT] \nonumber \\
&D_{\rm eff}/D=& \beta_\pm
\label{plecis} 
\end{eqnarray}
with $\bar \rho_{\pm} (x)$ the (local) ion concentration in the nano-channel, averaged over the channel width $h$; $q$ the ion charge;
$\rho(x)$ the local salt concentration in absence of electrostatic interactions (fixed by the salt concentration in the reservoirs); and $V(z)$ the electrostatic potential across the channel, for which Plecis \etal used a 
(linearized) Poisson-Boltzmann expression. 
A very good agreement with experimental results is found, as shown in Fig.~\ref{fig:Plecis}, showing that this exclusion-enrichement picture does capture the essential ingredients of the ion diffusive transport. 
Note that using the Donnan description above, one may furthermore approximate the electrostatic potential by its Donnan expression in Eq.(\ref{Donnan}). This
leads to an analytical expression for $\beta_\pm$ as
\begin{equation}
\beta_\pm=\left({\rho_\pm\over \rho_\mp}\right)^{1/2}
\end{equation}
where, as shown in Eq.(\ref{Donnan}), $\rho_\pm=\sqrt{\rho_s^2+\left({\Sigma\over h} \right)^2}\pm{\Sigma\over h}$. This expression reproduces the results in Fig.~\ref{plecis}.

Finally, one should note however that due to the difference in ion permeability, a charge separation will build up between the two ends of the channel. This leads therefore to the creation of a reacting electric field along the channel, which will compensate dynamically for this charge separation \cite{Plecis05}. This points to the complex couplings
associated with charge transport in nanochannels. More experimental and theoretical work is certainly in order to get further insight in these phenomena. 

\subsubsection{Pre-concentration} 

\begin{figure}[htb]
\begin{center}
\includegraphics [width=8 cm] {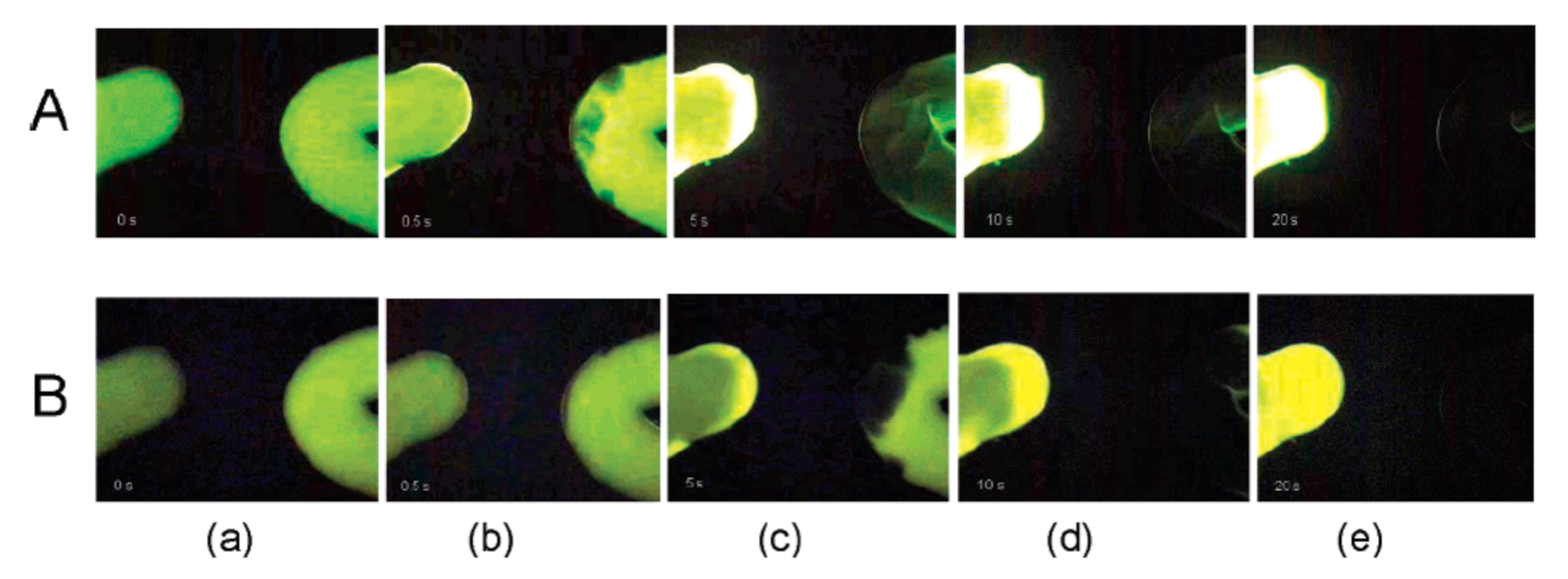}
\end{center}
\caption{ Two micro-channels filled with fluorescent probes are connnected by a nanochannel (a). After an electric field is applied across the nanochannels, an ion-enrichment and ion-depletion occurs at each end of the nanochannel (b-e).
Figures (a) to (e) show the evolution of the fluorescence in the microchannel versus time (from 0 to 20 s).
The probe used was (A) fluoresceine, (B) rhodamine 6G.
From Ref. \cite{Pu04}.}
\label{fig:Pu}
\end{figure}

Along the same lines,  another consequence of Debye layer overlap is the 'pre-concentration' phenomenon, which was first observed by Pu \etal \cite{Pu04}, and others \cite{Wang05}. Instead of a concentration drop as above, a voltage drop is applied along the nanochannel. 
It is then observed that ions enrich at one end and deplete at the other end. This phenomenon is illustrated in Fig.~\ref{fig:Pu}. 

As above, a key point underlying the phenomenon is that ions of the
opposite sign to the channel's surface, counter-ions, are more heavily transported than co-ions. This will create an enriched/depleted zone for both
ions at the two ends of the channel.  

This phenomenon has attracted a lot of interest, due in particular to its potential applications in the context of chemical analysis, for which it would provide a very interesing way of enhancing the sensitivity of detection methods.

At a basic level, the origin of preconcentration is a ``cross effect'' as introduced in Eq. (\ref{Mat-transport}): an electric potential drop $\Delta V$ leads to a
flux of ions ($\Phi_t$), as quantified by the cross-mobility $\mu_K$. According to the PNP result for $\mu_K$ ($\mu_K=e\,\mu  \times {2\SigmaÊ/h}$), 
this cross effect is thus a direct consequence of confinement and the existence of surface charges. 
Accordingly the magnitude of its effect, as compared \eg to the diffusive flux,  depends on the Dukhin number, see Eq.(\ref{mobility}).

But the detailed mechanims underlying the preconcentration involve non-linear couplings between the various transport process of ions in the nanochannel, which are  quite complex to rationalize.
A systematic description of the phenomenon was proposed by Plecis \etal in Ref.\cite{Plecis08}, revealing the existence of various preconcentration regimes.
We refer to Ref.\cite{Schoch08} for a detailed discussion on the mechanisms underlying this process.

\subsubsection{More complex functionalities: nanofluidic diodes}

The analogy with transport in semiconductors quoted above suggest that more complex nanofluid transport phenomena can be obtained
 in the regime of Debye layer overlap.

We discuss here the analog of a PN junction in SC transport. Using the equivalence table, Table I, a PN junction corresponds for
nanofluidic transport to a nanochannel exhibiting a disymetric surface charge along its surface.

\begin{figure}[htb]
\begin{center}
\includegraphics [width=4 cm] {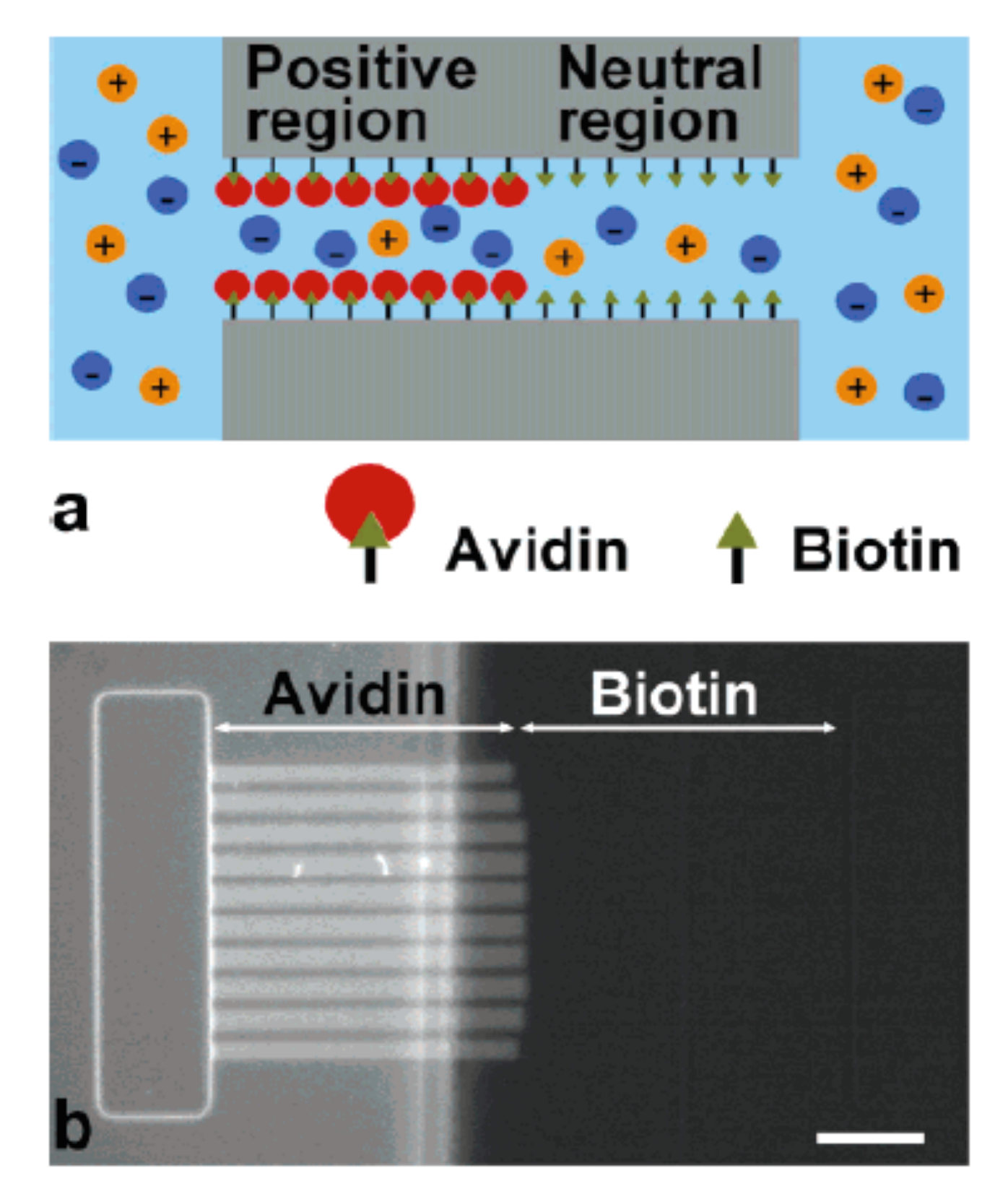}
\includegraphics [width=4 cm] {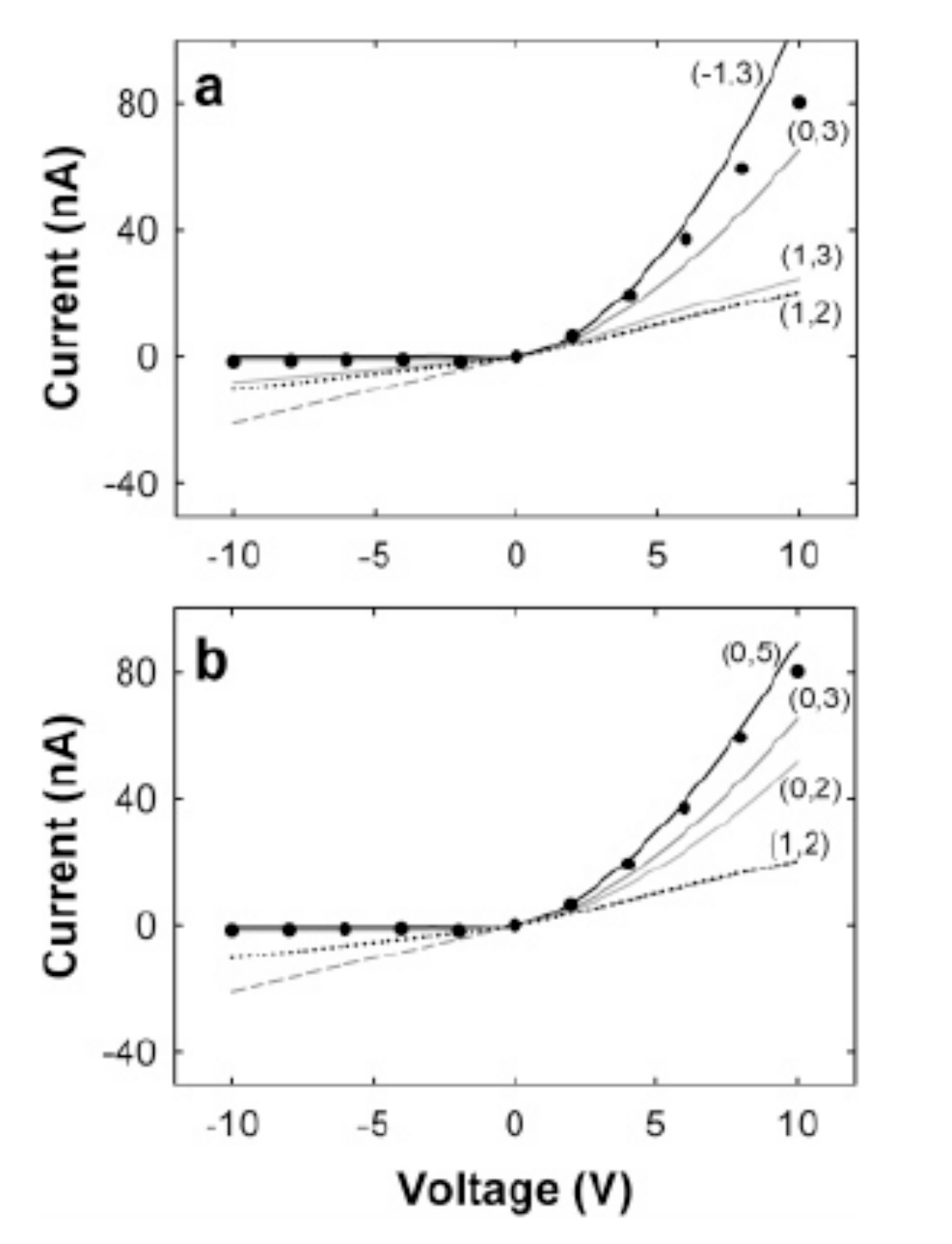}
\end{center}
\caption{ Ê{\it Left} (a) A nanofluidic diode is fabricated by patterning a nanochannel with different coatind on its
two moieties (avidin and biotin here). 
(b) 
Epifluorescence image of the fabricated nanofluidic diode, showing
fluorescently labeled avidin in half the channel. Scale bar 20 $\mu$m.
 {\it Right} current versus applied voltage. Symbols are experimental data, while the solid lines 
are theoretical predictions using a PNP transport model. From Ref. \cite{Karnik07}.}
\label{fig:Karnik}
\end{figure}

Such a nanofluidic diode device has been developped by Karnik \etal \cite{Karnik07}, following a previous work by Siwy \etal
in a different pore geometry \cite{Siwy02}.
This is illustrated in Fig.\ref{fig:Karnik} from Ref. \cite{Karnik07}. The surface of a nanochannel is coated with two different surface
treatments (half with avidin, half with biotin), leading to a surface charge contrast along the two moieties of the channel, Fig. \ref{fig:Karnik}-{\it Left}.
Accordingly  the current versus applied electric potential drop characteristics is found to 
exhibit an asymetric shape: as in a classical diode, the current passes only in one direction. Similar asymetric I-V curves are obtained for pores with asymetric
geometries, like conical pores obtained by track-etch techniques  \cite{Siwy02}.

At a more quantitative level, Karnik \etal discussed the effect on the basis of the PNP equations discussed above, under the assumption of local electroneutrality. 
In the present disymetric case, Karnik \etal solve numerically  these PNP equations  for an applied potential drop at the ends of the pore \cite{Karnik07}. Solutions are shown in Fig.~\ref{fig:Karnik}, demonstrating that the PNP framework is indeed able to capture the transport rectification measured experimentally.

This diode behavior can also be discussed in the context of the analogy with semiconductor transport that we put forward above.
The geometry described in Fig.~\ref{fig:Karnik} is indeed equivalent to that of a classical PN junction, corresponding to two regions with different doping. In the present case of a nanofluidic diode, the doping contrast between the P and N regions is associated with the different surface charge between the two
moieties of the nanochannel.
The diode effect in PN junctions is classically interpreted in terms of theÊ Space-Charge Zone at the interface between the two regions with different doping.
An approximate solution of the PNP equations can be proposed and leads to a rectified I-V characteristic 
which takes the form of the so-called  Shockley equation \cite{Kittel}
\begin{equation}
I=I_S\left( \exp\left[{{e V \over k_BT}}\right]-1\right)
\label{diode}
\end{equation}
where $I_S$ is the so-called
the saturation current $I_S$. 
This expression is expected to describe well the  blocking to non-blocking transition of the junction, \ie for moderate $eV/k_BT$.
However this expression strongly overestimates the current at large $eV/k_BT$ for which a full solution of the PNP
equations is needed. 

In the Shockley equation, Eq.(\ref{diode}), the saturation current is expressed in 
in terms of the donor-acceptorÊ densities, $\rho_A$ and $\rho_D$, as
\begin{equation}
I_S \approx e {\cal A}\, n_i^2\,  \left( {D_n \over L_n \rho_A} + {D_p\over L_p \rho_D} \right)
\end{equation}
where $n_i$ is the carriers density, $D_n$, $D_p$ the diffusion coefficients of electrons and holes, 
and $L_n$, $L_p$ the length of the P and N regions.
Using the equivalence table, Table I, one may convert this expression for the ionic transport
using $D_n=D_-$, $D_p=D_+$, $L_n=L_p=L/2$, while 
 $\rho_A=2\vert \Sigma_A\vert/h$ and $\rho_D=2\vert \Sigma_D\vert/h$.  

This Shockley expression predicts an I-V curve in  qualitative agreement with the experimental result in Fig.~\ref{fig:Karnik}. However, quantitatively it does (very strongly) overestimate the current for large voltages $V$, as expected due to the simplifying assumptions underlying the expression in Eq.~(\ref{diode}).

The analogy is however interesting to capture the underlying physics behing ion rectification in ionic transport.
It allows to predict a full zoology of fluidic functionalities, in line with their SC analogues.



Finally it is interesting to quote that the disymetry measured here for I-V characteristics generalizes to other transport phenomena. 
For example,  Siwy \etal also demonstrated the occurence of a disymetric diffusionÊ
of ions through conical pores \cite{Siwy05}.

\vskip1cm
Altogether these phenomena demonstrate the richness and complexity of behaviors obtained  in the regime of Debye layer overlap. This definitely shows the
great potential of nanofluidics, where further phenomena should emerge in the future taking benefit of these couplings.

\vskip1cm

\section{Thermal fluctuations}

In this section we raise the question of the role of thermal fluctuations in nanofluidic transport. 
As a general statistical rule, when the size of the system decreases, fluctuations play an increasingly important role.
But under which conditions do they play a role ? and which role could they play in nanofluidic transport ?

There are quite a few studies on this question in relation to nanofluidic transport. One may cite the experimental study of noise in solid state nanopores by
Dekker \etal \cite{Smeets08}. The question of thermal fluctuations was also discussed in the context of numerical molecular dynamics simulations
of osmotic flow through carbon nanotubes by Hummer \etal \cite{Berezhkovskii02,Kalra03}. Simulations show that -- in the single file regime -- 
the flow rate is essentially governed by thermal fluctuations rather than hydrodynamics. Hummer \etal proposed a 1D random walk to describe 
the nanofluidic water flow in this stochastic regime. 
Now, it is fair to remark that fluctuations are indeed expected to play a key role for single file transport, for which dynamics are highly correlated.

Aside from this situation, various indications of the role of thermal fluctuations have been reported, especially in studies involving capillary dynamics.
One may cite the noise effects in the breakup of fluid nanojets by Moseler and Landman \cite{Moseler00}, the noise assisted spreading of drops 
\cite{Davidovitch05}, and the influence of thermal noise in thin film dewetting \cite{Grun06} in relation to experiments \cite{Becker03a}. 
In these different cases, thermal noise modifies quite strongly the dynamics, even at relatively large length scales (up to typically 100nm in Ref.\cite{Grun06}).
These different situations can be described within the framework of fluctuating hydrodynamics \cite{Landau}, in which a random stress tensor
is added to the Navier-Stokes terms. While this approach is indeed fruitful in describing the role of thermal fluctuations, a general 
criterion to quantify the importance of noise is however lacking. 

Here, to illustrate the potential importance of fluctuations, we consider a simple situation in which an osmotic flow is driven across a single nanopore.
The pore has diameter $d$ and length $L$ and is {\it impermeable} to the solute. The pressure drop across the membrane is $\Delta p=k_BT \Delta c$, 
with $c$ the solute concentration.
Under the pressure drop $\Delta p$, a fluid flow is induced, with mean velocity $\bar u$. Assuming NS equations to hold gives
$\bar u \sim {d^2 \over \eta} {\Delta p \over L}$.  The flow is in the direction of higher solute concentrations.

Now one may ask the question: could fluctuations yield a reverse flow, \ie against the pressure drop, at least for a short time ? 
This would be clearly a (punctual) violation of second law. And what is the minimal size at which this may occur ?
A lead to such questions is provided by the so-called Fluctuation Theorem, which quantifies the probability of
second law violation by fluctuations. This domain is presently a very active, in particular in the context of single molecule spectroscopy \cite{Ritort06}. 
The theorem states that the probability $P(Q_\tau)$ to find a value $Q_\tau$ of the amount of heat dissipated in a time interval $\tau$
satisfies in a non-equilibrium state \cite{Evans93,vanZon03,vanZon04}
\begin{equation}
P(Q_\tau)/P(-Q_\tau)=\exp(Q_\tau/k_BT)
\label{FT}
\end{equation}
In the nanofluidic situation, this theorem would thus quantify the probability of a flow {\it against} the osmotic gradient. 
Let us compute the order of magnitude for the averaged $\bar Q_\tau$ in the stationary state: $\bar Q_\tau \approx \bar F \times \bar u \times \tau$,
with $F$ the frictional force acting on the nanopore surface. One has $F=\pi d L \sigma_w$ with $\sigma_w$ the
stress at the wall given by $\sigma_wÊ\sim d \Delta p/L$. To fix ideas, we compute the amount of heat dissipated over a time
needed for a molecule to pass through the whole pore, \ie $\tau = L/\bar u$. Altogether this gives
\begin{equation}
\bar Q_\tau \sim \Delta p \times {\cal V}
\end{equation}
with ${\cal V}={\pi \over 4} d^2 L$ the volume of the pore. 
According to Eq. (\ref{FT}), violations of the second principle are more likely to occur when $\bar Q_\tau$ is of the order of the
thermal energy.
Putting numbers, if we choose $\Delta c=1 M$ (close to physiological conditions), then the condition $Q_\tau \sim k_BT$ is
obtained for {\it nanometric} volumes, \ie for a pore with a size (diameter, length) typically in the nanometer range.
This is indeed typically the order of magnitude for the size of biological pores. 

The occurence of such 'counter flow' fluctuations has actually been observed in MD simulations of osmotic flows through carbon nanotube
membranes by Hummer \etal \cite{Kalra03}, and a stochastic model was proposed to account for the water transport across the tube. The above analysis thus 
fixes the limit where fluctuations starts to be predominant over the mean behavior. 

Finally we note that one may perform the same estimate for a different situation of an electric current induced by an electrostatic potential drop $\Delta V$.
Assuming a bulk conductance, the dissipated heat is estimated under quite similar terms as
$\bar Q_\tau \sim e\Delta V \times \rho_s {\cal V}$, with $\rho_s$ the salt concentration. Again the condition $Q_\tau \sim k_BT$
is achieved for {\it nanometric} volumes.

These results  point to the crucial role of fluctuations in pores with nanometric size. 
This is the typical scale of biological pore:
it is then interesting to 
point out that biological pores are working at the edge of second law violation ($\bar Q_\tau \sim k_BT$). 

These questions would deserve further experimental and theoretical investigations.

\section{Discussion and perspectives}

In this review, we hope  to have convincingly shown that nanoscales do indeed play a key  role in fluidics. 
A broad panel of length scales ranging from the molecular to the micrometric scales leads to 
a rich ensemble of nanofluidic phenomena. While the domain of validity
of Navier-Stokes equations was shown to extend down to the nanometer scale, specific transport
phenomena show up due the complex couplings which build up between flow and ionic transport, electrostatics, surface dynamics, etc.

We focused in this review on the behavior of fluids at the nanoscale. But in doing so we omitted
several important topics in the discussions, however strongly connected to nanofluidics questions.

This concern in particular the field of transport of (macro)molecules, polymers, polylelectrolytes or biological molecules (DNA, RNA) through nanopores. This involves either biological nanopores, like the widely studied 
$\alpha$-haemolysin, or artifical solid-state nanopores made by ionic drilling of membranes \cite{Dekker07}.
Starting from the pioneering work by Bezrukov \etal and Kasianowicz \etal \cite{Bezrukov93,Bezrukov96, Kasianowicz96}, 
there has been a thorough exploration of the translocation mechanisms of macromolecules in tiny pores \cite{Oukhaled07,Dorp09}, with important implications for the understanding of biological translocation process and potentially for single molecule analysis. This domain is now rapidly expanding and we refer to the recent review by C. Dekker for further reading \cite{Dekker07}.

Another aspect that we left aside is the question of nano- and micro- structured surfaces, in line with the recently
developped super-hydrophobic surfaces exhibiting the Lotus effect. Superhydrophobic coatings can lead to huge slippage effects, with slip lengths in the micrometer range  \cite{Cottin03,Ou05,Joseph06,Lee08}, as well as other dynamic phenomena, such as hydro-elastic couplings at the superhydrophobic interface \cite{Steinberger07}.
Such surfaces offer the possibility to considerably enhance the efficiency of transport phenomena in particular
in the context of slip enhancement discussed above. 
For example, as we pointed out in the text,  a massive amplification, 
by a factor up to $10^4$ (!), is predicted for diffusio-osmosis on superhydrophobic surfaces, which offers the possibility to device efficient salt pumps \cite{Huang08a}. The underlying coupling mechanisms remain however subtle, and this enhancement was shown for example to break down for electro-osmotic
transport \cite{Huang08a,Squires08}. The implications of nano- and micro- structuration on fluidic transport
is still at its infancy and remains to be thoroughly explored. It offers the possibility to couple
fluid dynamics over the scales, from the nano- to the micro-scales, and even to larger macroscopic scales
\cite{Duez07}.


Furthermore, beyond these questions, it is interesting to note that nanofluidics offer the possibility to attack old 
questions with new point of views, and new control possibilities. 
We quoted the amazing flow permeability measured by carbon nanotube membranes. 
While this performance remains to be understood -- and possibly investigated at the individual
nanotube level --, this opens very promising application in the 
field of energy, in particular for portable energy sources. Such membranes indeed offers the possibility of a very efficient 
conversion of hydrostatic energy to electrical power at small scales \cite{Pennathur07,Ren08}, see also \cite{Zhao08}.
As we have discussed above,  low (nanoscale) 
friction at surfaces considerably amplifies electrokinetic processes, such as 
electro-osmosis or streaming currents \cite{Joly04,Bouzigues08}. The latter phenomenon is able to produce 
an electrical current from a pressure drop \cite{vanderHeyden06,vanderHeyden07}. On this basis it 
was argued \cite{Pennathur07, Ren08} that the efficiency of this mechano-electric 
transduction raises considerably for slippery walls : reasonable values suggest an 
increase of the energy conversion efficiency from a few percents to 30\% for 
slippery surfaces ! While the tremendous effect of low surface friction and 
slippage on charge transport has been recently confirmed experimentally 
\cite{Bouzigues08}, its expected impact on mechano-electric energy 
conversion has not received an experimental confirmation.  
Extrapolating with the results obtained with the nanotube membranes, this 
power conversion efficiency raises to close to 100\% (due to the expected 
extremely weak dissipation) and this amazing prediction would lead to a 
production of electrical power of several kW/m2  for a pressure drop of 
1bar \cite{Ren08}: this is an impressive result, which has not been confirmed up to now. 
While such predictions have to be taken with care, they strongly suggest to 
explore the role of nanofluidics in the context of energy conversion. Such 
nanofluidic energy conversion devices have a priori the potential to power larger 
scale systems (for example in cars or portable devices), due to their high power 
density and low weight \cite{Eijkel05}.

An alternative field of application for the progress made in nanofluidic transport is desalination.
Indeed, as half of the humanity has no immediate access to potable water,  desalination 
of seawater has emerged over the recent years as an alternative solution to 
develop fresh water. 
Reverse osmosis is one of the techniques used in this context, for example in the Ashkelon plant in Israel \cite{Yermiyahu07}: 
seawater is pushed through a membrane impermeable to 
the salt. Such a process is expensive (typically around 0.5\$/m$^3$ for the 
Ashkelon plant in Israel \cite{Yermiyahu07}),  due -- among other factors -- to the 
energy required to this operation. The possibility of reducing this energy by using 
a considerably more permeable membrane should have therefore a direct impact
on the cost of produced water. This is a potential application of membranes 
made of carbon nanotubes which, as we discussed
\cite{Majumder05,Holt06,Whitby08}, exhibit a permeability 2 to 3 orders of magnitude higher 
than those with micropores, a fully unexpected and still debated result.  
Another question for the desalinated water is the mineral composition of the 
produced water which -- if not controlled -- could have a deep, negative, impact 
on agriculture and health. For example the concentration of Boron (B), which is 
toxic to many crops, is high in seawater, and requires to be partially eliminated 
by desalination post-treatments \cite{Yermiyahu07}. Reversely, desalination removes 
some ions which are essential to plant growth, such as Mg$^{2+}$ or SO$_4^{2-}$. There is 
therefore a potential need for a better selectivity of filtration. 

Achieving 
selectivity similar to biological channels, such as aquaporins 
\cite{Walz94,Sui01,Borgnia01}, is still out of reach, but nanofluidics has a major role 
to play in this context in order to find the key tuning parameters and 
architecture. 
Such applications may still be years away, but the basic 
principles underlying their operation will develop in the years to come.

\begin{acknowledgments}
We thank A. Ajdari, L. Auvray, C. Barentin, J.-L. Barrat, C. Cottin-Bizonne, R. Netz, D. Stein,
and C. Ybert
 for many stimulating discussions. LB thank in particular D. Stein for illuminating discussions on the conductance
 effects.

\end{acknowledgments}


\end{document}